\documentclass[aps,prl twocolum,amsmath,amssymb,superscriptaddress,nofootinbib]{revtex4-2}
\usepackage{graphicx}
\usepackage{amsmath}
\usepackage{color}
\usepackage{bm}
\usepackage{amssymb}
\usepackage{mathrsfs}
\usepackage{amsbsy}
\usepackage{subfigure}
\newcommand{\meqref}[1]{Eq.\eqref{#1}}
\newcommand{\mpref}[1]{Figure \ref{#1}}
\newcommand{\be}{\begin{equation}}
\newcommand{\ee}{\end{equation}}
\newcommand{\bea}{\begin{eqnarray}}
\newcommand{\eea}{\end{eqnarray}}
\begin{document}

\title{Chaos of Particle Motion near the Black Hole with Quasi-topological Electromagnetism}
	
\author{Yu-Qi Lei}
\affiliation{Department of Physics, Shanghai University, Shanghai 200444, China}
\author{Xian-Hui Ge}
\email{gexh@shu.edu.cn}
\affiliation{Department of Physics, Shanghai University, Shanghai 200444, China}
\affiliation{
Center for Gravitation and Cosmology, College of Physical Science and Technology,
Yangzhou University, Yangzhou 225009, China}
\author{Cheng Ran}
\affiliation{Department of Physics, Shanghai University, Shanghai 200444, China}
\begin{abstract}
We explore the chaotic behavior of particle motion in a black hole with quasi-topological electromagnetism. The chaos bound is found to be violated in the higher order expansion of the metric function and the
electric potential near the horizon. We draw the Poincar\'e sections of particle motion corresponding to the chaos bound violated and non-violated cases, respectively. Then we study the relationship between the ``maximal'' Lyapunov exponent $\lambda_s$ defined by the static equilibrium and the Lyapunov exponent of the particle geodesic motion near the Reissner-Nordstr$\rm\ddot{o}$m(RN) black hole and the black hole with quasi-topological electromagnetism. We find an interesting relationship between the Lyapunov exponent $\lambda_{ph}$ of photon's radial falling into the black hole and the ``maximal'' Lyapunov exponent $\lambda_s$. For the black holes whose metric function increases monotonically with radius outside horizon, this leads to $\lambda_{ph} \geq 2\lambda_s$.
\end{abstract}

\maketitle

\section{Introduction }
Chaos is an important nonlinear phenomenon, which describes the violent response of the dynamic system to perturbations. Chaos of particle motion near the black hole has been studied for a long time, such as using the Melnikov method to find the chaotic orbit \cite{Bombelli:1991eg}, distinguishing the chaotic orbits of particles or strings from periodic orbits \cite{dettmann1995chaos,Dalui2018PresenceOH,Suzuki_199,letelier1997chaos,aless1999chaos,dalui2019induction,Kao:2004qs,Kopacek:2014gza,Chen:2016tmr,Wang:2016wcj,Ma:2014aha,Basu:2011di,Giataganas:2017guj} and probing the instability of particles motion \cite{cardoso2008geodesic,pradhan2012stability,pradhan2012isco,pradhan2013lyapunov,pradhan2014circular}. Recently, much attention has been paid to the quantum chaos originated from the AdS/CFT correspondence\cite{Maldacena2016ABO}. In 2015, Maldacena, Shenker and Stanford pointed out in quantum field theory that for a quantum system like a black hole, its Lyapunov exponent $\lambda$ should have an upper bound \cite{Maldacena2016ABO}
\bea
\lambda \leq 2\pi T,\label{chaos bound1}
\eea
where $T$ is the Hawking temperature of the black hole in the natural unit $\hbar=1$. Note that the Hawking temperature $T$ is related to the surface gravity $\kappa$ via $\kappa=2 \pi T$. In turn, the chaos bound can be recast as $\lambda \leq \kappa$. The chaos bound is closely related to the chaotic behavior in the strongly correlated system and the physics of black holes.
\\
 \indent The chaos bound can also be studied through examining the geodesic motion of probing particles near black holes. Susskind proposed that for a neutral particle falling radially toward a black hole, its Rindler momentum will increase exponentially with the exponent equaling to the surface gravity $\kappa$ \cite{Susskind2018WhyDT}, indicating the chaos bound is saturated. On the other hand, when a particle in its static equilibrium near the horizon with an external potential, the ``maximal'' Lyapunov exponent $\lambda_s$ can be obtained \cite{Hashimoto2017UniversalityIC} with the relation $\lambda_s = \kappa$ at the horizon. However, the chaos bound could be violated for charged particles outside charged black holes \cite{Zhao2018StaticEO} by studying the near-horizon expansion.
 \\
\indent Our motivations of this paper are to study the Poincar\'e section and the chaos of particle motion for a very special black hole solution with quasi-topological electromagnetism obtained in \cite{Liu2019QuasitopologicalED}. We hope to study the Lyapunov exponent of particle motion near this black hole to understand the chaos bound and black hole chaos more. Interestingly, we find that the chaos bound can be violated for this special black hole. And in the system where chaos bound can be violated, the chaos is indeed strengthened.
\\
\indent Subsequently, we calculate the Lyapunov exponent $\lambda_c$ of the circular geodesic motion of natural particles near black hole, and find that $\lambda_c$ is not constrained by the ``maximal'' Lyapunov exponent $\lambda_s$. Meanwhile, we are inspired by \cite{brown2018falling,ageev2018things,Zhou2018ParticleMA} in which the authors pointed out that as the speed of particle cannot reach the speed of light, the growth rate of the particle Rindler momentum will not saturate the chaos bound, or even increase exponentially. There may be a connection between particles' radial falling, the speed of light and chaos bound. So we study the radial falling of particles and find an interesting relation between the Lyapunov exponent $\lambda_{ph}$ of photon's radial falling and the ``maximal'' Lyapunov exponent $\lambda_s$. For the black hole whose metric function increases monotonically outside horizon, we can see there is $\lambda_{ph} \geq 2\lambda_s$.
\\
\indent This paper is organized as follows. In Section \ref{black hole chaos}, we briefly review the black hole chaos and motivate why the study of black hole chaos is interesting. In Section \ref{A special dyonic black hole}, we provide some information about the black hole with quasi-topolopical electromagnetism \cite{Liu2019QuasitopologicalED}. In Section \ref{Static equilibrium of charged particles}, we discuss the static equilibrium of charged particles near the special dyonic black holes to check whether the chaos bound $\lambda \leq \kappa$ will be violated, and perform a numerical analysis of a toy model. In Section \ref{Geodesic motion of neutral particles}, we study the geodesic motion of neutral particles, including circular geodesic motion and radially falling. In Section \ref{Conclusion and discussions}, we briefly summarize the main results. In Appendix \ref{FLI}, the static equilibrium of a charged particle outside the RN black hole is studied by the fast Lyapunov indicator. In Appendix \ref{Static equilibrium with only gravity}, we discuss the static equilibrium with only gravity. In Appendix \ref{sejm}, we calculate the ``maximal'' Lyapunov exponent $\lambda_s$ by the Jacobian matrix method. In Appendix \ref{upper bound}, we consider some details about the ``maximal'' Lyapunov exponent $\lambda_s$ outside these black holes. In Section \ref{toy model Parameters}, we discuss in detail the parameters for the Poincar\'e section in a toy model.

\section{Black hole chaos: a brief review}\label{black hole chaos}
Gravity theory is a nonlinear theory, and chaos should naturally exist in it. As the black hole is an important research object in gravity theory, the chaos near the black hole has high research value. Especially after the AdS/CFT duality is proposed, studying the connection between black hole theory and quantum chaos can help us further understand gravity theory and the essence of chaos. In this section, we briefly review and disscuss black hole and the related issues.
\\
\subsection{Some indicators for chaos around black holes}
In the early studies on black hole chaos, searching for the behavior of chaos is the main task. The non-integrability of dynamical equation in black holes was studied by some analysis methods \cite{Bombelli:1991eg}. Some indicators such as the Lyapunov exponent, the fast Lyapunov indicator and the Poincar\'e section were also proposed to check the existence of chaos near the black hole.
\subsubsection{Lyapunov exponent}
The Lyapunov exponent describes the average rates of expansion and contraction of two adjacent orbits in the classical phase space, which can be defined by
\be
\lambda=\lim_{t \to \infty}\left(\frac{1}{t} \right)\ln \left(\frac{\triangle x(t)}{\triangle x(0)} \right),
\ee
where $\triangle x$ is the distance between the two orbits. The Lyapunov exponent can be used to describe the perturbation's exponential increasion in chaotic motion. The positive Lyapunov exponent means the existence of chaos.
\\
\indent For the equilibrium of particles outside the black hole, the Lyapunov exponent can be obtained by the Jacobian matrix \cite{cardoso2008geodesic,pradhan2012stability,pradhan2012isco,pradhan2013lyapunov,pradhan2014circular}. The equation of motion of particles can be schematically written as
\be
\frac{dX_i}{dt}=F_i(X^j),
\ee
$X^i$ is the coordinates and $F_i(X^j)$ is a function to be determined. Considering a certain orbit, we can linearize the equation of perturbation
\be
\frac{d \delta X_i(t)}{dt}=K_{ij}(t)\delta X_j(t),\label{jmd}
\ee
where
\be
K_{ij}(t)=\left.\frac{\partial F_i}{\partial X_j}\right|_{X_i(t)}
\ee
is the Jacobian matrix. When the equilibrium of particles outside black holes is considered, the Lyapunov exponent can be given by $\lambda=\det(K_{ij})$.

\subsubsection{The fast Lyapunov indicator (FLI)}
The fast Lyapunov indicator (FLI) is a more effective tool to search for chaos \cite{FROESCHLE1997881}. FLI is usually defined by considering the difference between two nearby trajectories of particles, which is the so-called two-particle method or two-nearby-trajectory method. For chaotic motion, even for weak chaotic case, FLI will grow at an exponential ratio. In general relativity, $FLI(\tau)$ can be described as \cite{Wu:2003pe,Wu:2006rx}
\be
FLI(\tau)=-k[1+\log _{10} d(0)]+\log_{10} \left|\frac{d(\tau)}{d(0)} \right|,
\label{FLIeq}
\ee
where $d(\tau)=\sqrt{\left|g_{\mu \nu} \triangle x^\mu \triangle x^\nu\right|}$, $\triangle x^\mu$ is the deviation vector between two nearby trajectories at proper time $\tau$. To avoid two orbits expand too fast, the sequential number $k$ of renormalization is considered (the value of $k$ can take 0,1,2...). In this paper, we use FLI to analyse the perturbation's growth in the static equilibrium of charged particles in Appendix \ref{FLI}.

\subsubsection{Poincar\'e section}
The Poincar\'e section is the most commonly used tool to analyze dynamical systems, which automatically follows the non-integrability analysis in examining the existence of chaos. It can be defined as the intersection of a given hypersurface and motion trajectory in high-dimensional phase space (d$\geq$3)
. Using the Poincar\'e section, the periodic orbits, quasi-periodic orbits and chaotic orbits can be clearly distinguished in dynamic systems.
\\
\indent Using these indicators, many chaotic phenomena in black holes have been studied, such as the motion of spining particle \cite{Kao:2004qs}, scalar particle \cite{Wang:2016wcj}, and even strings \cite{Ma:2014aha, Basu:2011di,Giataganas:2017guj} and the chaos in the background of a massive magnetic dipole \cite{Kopacek:2014gza}, rotating black holes \cite{letelier1997chaos,Chen:2016tmr}, the black holes with halo \cite{letelier1997chaos,aless1999chaos}. However, there were only the chaotic phenomenon near the black hole shown. The essential characteristics of black hole chaos are still to be discovered.\\

\subsection{More on black hole chaos}
Black hole chaos is an important subject in the study of black hole theory, which can be related to many physical problems. Actually, there are many physical problems related to black hole chaos. There are many phenomena related to black hole observation that have attracted much attention, such as photon sphere \cite{Claudel:2000yi}, shadow and gravitational lensing \cite{Wang:2017qhh}. It is natural to discuss black hole chaos in these problems. The quasinormal modes (QNMs) of black holes describe the damping and oscillation of the gravitational waves emitted by the perturbation of black hole \cite{Kokkotas:1999bd,ching1994quasinormal,ching1995wave,nollert1996significance,horowitz1999quasinormal,dolan2009expansion,yang2012quasinormalmode}. In \cite{cardoso2008geodesic,pradhan2012stability,pradhan2012isco,pradhan2013lyapunov,pradhan2014circular}, the Lyapunov exponent of circular null geodesic motion was related to QNMs, which gives an easier method to calculate the QNMs.
\\
\indent From the momentum-size duality proposed in \cite{Susskind2018WhyDT}, we can see that the black hole chaos is a bridge between gravity and quantum theory. Follow this duality, the connections between momentum and quantum complexity were explored in \cite{susskind2019complexity,mohapatra2019sizemomentum,barbn2019momentumcomplexity,colangelo2020chaos}, where the quantum complexity \cite{susskind2014addendum} is a very meaningful concept that can be related to many physical topics such as the action of black holes \cite{brown2015complexity,brown2015}, the shock wave geometry \cite{stanford2014complexity}, the accelerated expansion of the universe \cite{ge2017quantumy} and the partition function \cite{sun2019complexity}. There are also many more problems related to black hole chaos, such as black hole thermodynamics \cite{dalui2019horizon,dalui2020near}, acoustic black holes \cite{wang2019geometry} and gauge/gravity correspondence \cite{Zayas:2010fs}. Therefore, it is worthwhile to study the chaos of particles near the black hole with quasi-topological electromagnetism.

\section{A special dyonic black hole}\label{A special dyonic black hole}
Let us consider a black hole with quasi-topological electromagnetism proposed by \cite{Liu2019QuasitopologicalED}. The authors in \cite{Liu2019QuasitopologicalED} considered a 4-dimensional gravitational theory including the pure cosmological constant $\Lambda_0$ and the minimum coupling electromagnetic interaction. Its Lagrangian is given by \cite{Liu2019QuasitopologicalED}
\bea
\mathcal{L}=\sqrt{-g}\left[R-2\Lambda_0 -\alpha_1 F^2-\alpha_2\left(\left(F^2\right)^2-2F^{(4)} \right) \right]\label{Dl},
\eea
where $\alpha_1$ and $\alpha_2$ are coupling constants and $F^2=F^{\mu \nu}F_{\mu \nu}$, $F^{(2)}=F_{\nu}^{\mu}F_{\mu}^{\nu}=-F^2$, $F^{(4)}=F^{\mu}_{\nu}F^{\nu}_{\rho}F^{\rho}_{\sigma}F^{\sigma}_{\mu}$. The corresponding Maxwell's field equation is
\bea
&\nabla _{\mu}\tilde{F}^{\mu \nu}=0\label{DM},
\eea
where $\tilde{F}^{\mu \nu}=4\alpha_1 F^{\mu \nu}+8\alpha_2 \left( F^2 F^{\mu \nu}-2F^{\mu \rho}F^{\sigma}_{\ \rho}F_{\sigma}^{\ \nu}\right)$. The Einstein's field equation can be written as
\bea
R_{\mu \nu}-\frac{1}{2}Rg_{\mu \nu}+\Lambda_0 g_{\mu \nu}=T_{\mu \nu}\label{DE}.
\eea
This theory yields a static dyonic black hole solution
\bea
ds^2=-f\left( r\right)dt^2+\frac{dr^2}{f\left( r\right)}+r^2d\Omega^2_{2,\epsilon}\label{metric},
\eea
where the metric $d\Omega^2_{2,\epsilon}$ corresponds to a two-dimensional hyperbolic, torus and sphere, respectively with $\epsilon$ takes values -1, 0, 1. The metric function can be expressed as \cite{Liu2019QuasitopologicalED}
\bea
f\left(r \right)=-\frac{1}{3}\Lambda_0r^2+\epsilon -\frac{2M}{r}+\frac{\alpha_1p^2}{r^2}+\frac{q^2}{\alpha_1 r^2}\,_2F_1\left(\frac{1}{4},1;\frac{5}{4};-\frac{4p^2\alpha_2}{r^4\alpha_1} \right)\label{Dmf},
\eea
where $M$ is the mass of black hole, $\Lambda_0$ is the cosmological constant, $q$ is the electric charge, $p$ is the magnetic charge and $\,_2F_1\left(\frac{1}{4},1;\frac{5}{4};-\frac{4p^2\alpha_2}{r^4\alpha_1} \right)$ is a hypergeometric function.
\\
\indent The electric and magnetic charges can be obtained from electromagnetic tensor $\tilde{F}_{\mu \nu}$ which is defined by Maxwell's field equation \meqref{DM}
\bea
Q_e=\frac{1}{4\pi}\int \tilde{F}^{0r}=q,\qquad Q_m=\frac{1}{4\alpha_1 \pi}\int F=\frac{p}{\alpha_1}\label{charge}.
\eea
So the electric and magnetic potential functions are given by \cite{Liu2019QuasitopologicalED}
\bea
\begin{aligned}
\Phi_e \left( r\right)&=\frac{q\,_2F_1\left(\frac{1}{4},1;\frac{5}{4};-\frac{4p^2\alpha_2}{r^4\alpha_1} \right)}{\alpha_1r},
\\
\Phi_m \left( r\right)&=-\frac{q^2\,_2F_1\left(\frac{1}{4},1;\frac{5}{4};-\frac{4p^2\alpha_2}{r^4\alpha_1}\right)}{4pr}+\frac{\alpha_1 q^2r^3}{4p\left(4\alpha_2p^2+\alpha_1r^4\right)}+\frac{\alpha_1^2p}{r}.
\end{aligned}\label{Dpf}
\eea
When the constant parameters ($M,p,q,\Lambda_0,\alpha_1 , \alpha_2$) take proper values, different types of black hole solutions can be obtained.
\\
\indent Set $\alpha_1=1, \Lambda_0=0, \epsilon=1$, two types of dyonic black holes can be obtained as the constants $(\alpha_2, q, p)$ in the general solution \meqref{Dmf} take appropriate values. Note that in these black hole solutions, as $M$ takes different values, the black hole will also have different properties.
\subsubsection*{$\bullet$ Case 1: black holes with at most two horizons}
Consider a special condition of \meqref{Dmf} with the parameters
\begin{equation*}
(q^2,p^2,\alpha_2)=(\frac{5}{2},\frac{1}{2},2).
\end{equation*}
The corresponding function and electric potential functions become
\bea
\begin{aligned}
f\left(r\right)&=1-\frac{2M}{r}+\frac{1}{2r^2}\left(1+5_2F_1[\frac{1}{4},1;\frac{5}{4};-\frac{4}{r^4}]\right),
\\
\Phi_e \left(r\right)&=\frac{1}{2r}\left(1+5_2F_1[\frac{1}{4},1;\frac{5}{4};-\frac{4}{r^4}] \right).
\end{aligned}\label{case1}
\eea
It is one of the first type black hole solutions with a constant $M_0$ (its value is related to the value of $q$, $p$, $\alpha_2$ and the concrete definition of $M_0$ can be found in \cite{Liu2019QuasitopologicalED}). When the black hole mass M takes different values, different situations can be found
\begin{equation*}
\left\{
\begin{aligned}
M>M_0 \qquad &black\ holes\ with\ two\ horizons,
\\
M=M_0 \qquad &extremal\ black\ hole,
\\
M<M_0 \qquad &there\ is\ a\ naked\ singularity\ at\ r = 0.
\end{aligned}
\right.
\end{equation*}
For \meqref{case1}, the corresponding constant $M_0=1.6372384$. In the subsequent calculation process, for the first type of black hole solutions, we will consider an example with two horizons. In particular, $M=1.996$, two horizons locate at
\begin{equation*}
r_-=0.28617, \qquad r_+=3.00002.
\end{equation*}
For this black hole, the metric function $f\left(r\right)$ is monotonically increasing outside the outer horizon and the rate $\frac{df\left(r\right)}{dr}$ is monotonically decreasing.
\subsubsection*{$\bullet$ Case 2: black holes with at most four horizons}
There is another special case for \meqref{Dmf} with parameters:
\begin{equation*}
(q^2,p^2,\alpha_2)=(\frac{20868}{443},\frac{396}{443},\frac{196249}{1584}).
\end{equation*}
The corresponding metric and electric potential functions can be given by
\bea
\begin{aligned}
f(r)&=1-\frac{2M}{r}+\frac{12}{443r^2}\bigg(33+1739_2F_1[\frac{1}{4},1;\frac{5}{4};-\frac{443}{r^4}]\bigg),
\\
\Phi_e(r)&=\frac{12}{443r}\bigg(33+1739_2F_1[\frac{1}{4},1;\frac{5}{4};-\frac{443}{r^4}]\bigg).
\end{aligned}\label{case2}
\eea
The second type black hole solutions with three constants: $M_-, M_+, M_0$ (their values depend on the values of $q$, $p$, $\alpha_2$ and the concrete definition of them can be found in \cite{Liu2019QuasitopologicalED}). Similarly, different situations depend on the value of M:
\begin{equation*}
\left\{
\begin{aligned}
M<M_-  	  \quad &There\ is\ no\ horizon\ here\ but\ a\ naked\ singularity\ at\ r=0,
\\
M_-<M<M_+ \quad &black\ holes\ with\ two\ horizons\ and\ the \ equilibrium\ of\ Newton's\ potential,
\\
M_+<M<M_0 \quad &black\ holes\ with\ four\ horizons,
\\
M>M_0 \quad &black\ holes\ with\ two\ horizons,
\end{aligned}
\right.
\end{equation*}
where the corresponding $M _-, M _+$, and $M_0$ are given by
\begin{equation*}
M_-=6.6316,\quad M_+=6.7730,\quad M_0=6.9135.
\end{equation*}
In the subsequent calculations, we will choose the values of M = 6.7, 6.8, 7.0 as concrete examples.
\begin{flushleft}
\textbf{Case 2-1: Black hole with Newtonian potential equilibrium}
\end{flushleft}
\begin{equation*}
M=6.7, \qquad \qquad r_-=0.66823,\quad r_+=1.54979.
\end{equation*}
For this black hole, the metric function $f\left(r\right)$ is not monotonically increasing outside the outer horizon, and the Newton potential has equilibriums at $r=2.8480$ and $r=6.1421$. So it is possible for particles to maintain a static equilibrium at these positions without any external forces, where the Lyapunov exponent is discussed in Appendix \ref{Static equilibrium with only gravity}.
\begin{flushleft}
\textbf{Case 2-2: Black hole with cosmological horizons}
\end{flushleft}
\begin{equation*}
M=6.8,\qquad \qquad r_-=0.53736,\quad r_+=2.07702 ,\quad r_1=5.16615, \quad r_2=6.79477,
\end{equation*}
where $r_-$, $r_+$ are the two horizons of this black hole, and $r_1$, $r_2$ are the so-called cosmological horizons. The metric function $f\left(r\right)$ increases firstly and then decreases between the outer event horizon and the inner cosmology horizon. This black hole also has an equilibrium of the Newton potential at $r=3.06723$.
\begin{flushleft}
\textbf{Case 2-3: Black hole with two horizons and non-monotonic metric function increase rate}
\end{flushleft}
\begin{equation*}
M=7.0,\qquad \qquad r_-=0.40827,\quad r_+=8.34842.
\end{equation*}
 For this black hole, the metric function $f\left(r\right)$ is monotonically increasing outside the outer horizon, but the rate $\frac{df\left(r\right)}{dr}$ increases first and then decreases outside the outer horizon.
 \\
 \\
 \indent Actually, when the parameters of black holes satisfy different constraint equations, the black holes can be classified into these different types. For the same type of black holes, they have same properties. It is reasonable to follow the parameter values in \cite{Liu2019QuasitopologicalED}.

\section{Static equilibrium of charged particles outside the black hole}\label{Static equilibrium of charged particles}
In this section, we consider the general discussion of charged particles outside the black hole and derive the Lyapunov exponent. We consider a 4-dimensional static spherically symmetric black hole
\bea
ds^2=-f\left(r\right)dt^2+\frac{dr^2}{f\left(r\right)}+d\Omega^2 \label{smetric}.
\eea
At the horizon, the surface gravity $\kappa$ is
\bea
\kappa=\left.-\frac{1}{\sqrt{g_{rr}}}\frac{d\sqrt{g_{tt}}}{dr}\right|_{horizon}=\left.\frac{1}{2}f^{'}\left(r \right)\right|_{horizon}\label{kappa},
\eea
where the prime `` $'$ '' denotes derivative with respect to $r$.
\\
\indent The Lagrangian of a particle near this black hole can be written as\footnote{There is another form of Lagrangian $L=m\left(\frac{1}{2}g_{\mu \nu}\dot{x}^{\mu}\dot{x}^\nu +V(r)\right)$, which is often used to calculate particle motion. In the Appendix \ref{sejm}, we use the Jacobian matrix method to calculate the ``maximal'' Lyapunov exponent $\lambda_s$ in these two forms of Lagrangian, and the results are the same as the expression of $\lambda_s$ in \cite{Hashimoto2017UniversalityIC,Zhao2018StaticEO}}
\bea
\mathcal{L}=-m\left(\sqrt{-g_{\mu \nu}\dot{x}^{\mu}\dot{x}^{\nu}}+V\left(r \right)\right)\label{sl1},
\eea
where $m$ is the mass of the particle and $V\left(r \right)$ is the external potential in the radial direction. When $V^{'}\left(r \right)<0$, the potential $V\left(r \right)$ can provide the particle with a repulsive force away from the black hole, so that the particle may not fall into the black hole and maintain static equilibrium near the horizon. In this section, we take the static gauge $\tau =t$, so the dot ``$\cdot$'' denotes derivative with respect to the proper time `` $t$ '' in this section.
\\
\indent When the particle stays in its static equilibrium near the horizon, the particle's Lagrangian \meqref{sl1} describing its radial motion can be reduced to
\bea
\mathcal{L}=-m\left(\sqrt{f\left(r \right)-\frac{\dot{r}^2}{f\left(r \right)}}+V\left(r \right)\right)\label{sl2}.
\eea
Note that when the particle maintains static equilibrium, we have $\dot{r}\ll 1$. After expanding the Lagrangian in terms of $\dot{r}$, we can obtain the effective Lagrangian
\bea
\mathcal{L}_{eff}=\frac{\dot{r}^2}{2f\left(r \right)\sqrt{f\left(r \right)}} -V_{eff}\left(r \right),\qquad V_{eff}\left(r\right)=\sqrt{f\left(r \right)}+V\left(r \right)\label{effl1},
\eea
where $V_{eff}$ is the effective potential and at the particle's equilibrium position there is $V_{eff}^{'}=0$. After expanding $V_{eff}$ at the particle's static equilibrium position $r=r_0$, we have the effective Lagrangian satisfying the relation
\bea
\mathcal{L}_{eff}\sim \frac{1}{2f\left(r \right)\sqrt{f\left(r \right)}}\left[\dot{r}^2+\lambda^2 \left(r-r_0 \right)^2 \right]\label{effl2},
\eea
where $\lambda$ is the Lyapunov exponent obeying
\bea
\lambda^2=-f\left(r \right)\sqrt{f\left(r \right)}\left[\left(\sqrt{f\left(r \right)} \right)^{''}+V^{''}\left(r \right) \right]\label{slambda}.
\eea
In the equilibrium position $r=r_0$, $V_{eff}^{'}$=0. If $V_{eff}$ has the maximum $V_{eff}^{''}<0$, the static equilibrium is unstable and $\lambda^2>0$. If $V_{eff}$ has the minimum as $V_{eff}^{''}>0$, the static equilibrium is stable and $\lambda^2<0$.
\\
\indent From the effective Lagrangian \meqref{effl2}, we can see that the particle should follow the equation of motion at the equilibrium position
\bea
\ddot{r}=\lambda^2 \left(r-r_0 \right).
\eea
This equation of motion tells us the particle's trajectory in the radial direction, that is to say
\bea
r=r_0+Ae^{\lambda t}+Be^{-\lambda t}.
\eea
If $\lambda$ is real, the exponential increasion of $r$ may imply the existence of chaos. Actually, when there is no perturbation in other directions, the position $r=r_0$ (where $V_{eff}$ has the maximum) will be a separatrix of the phase space, which can provide a ``maximal'' Lyapunov exponent here \cite{levin2008homoclinic,perezgiz2008homoclinic,hackmann2009analytic,Hashimoto2017UniversalityIC}. When the external potential is strong enough, the position where $V_{eff}$ has the maximum can be near the horizon.
\subsection{Chaos bound at the horizon}
Returning to the black hole solution \meqref{Dmf}, we calculate the Lyapunov exponent when a charged particle maintains its static equilibrium near the horizon of these black holes and further verify the chaos bound $\lambda \leq \kappa$. Here, the external potential we are considering is provided by the electric field, and its form is
\bea
V\left(r \right)=\frac{e}{m}\Phi_e\left(r \right),
\eea
where $e$ and $m$ are the charge and mass of the particle, and $\Phi_e\left(r \right)$ is the electric potential function.
\\
\indent For the particle's static equilibrium at $r=r_0$, the Lyapunov exponent $\lambda$ satisfies the expression \cite{Zhao2018StaticEO}
\bea
\lambda^2=\left.f\left(r \right)\sqrt{f\left(r \right)}\left(\frac{\Phi_e^{''}\left(r \right)}{\Phi_e^{'}\left(r \right)}\left(\sqrt{f\left(r \right)} \right)^{'}-\left(\sqrt{f\left(r \right)} \right)^{''} \right)\right|_{r=r_0}\label{selambda}.
\eea
When the static equilibrium position is infinitely close to the black hole outer horizon $r=r_+$, the Lyapunov exponent $\lambda$ should have an upper bound, that is, the chaos bound $\lambda \leq \kappa$. The content of this section is mainly to calculate the Lyapunov exponent $\lambda$ of the charged particle near the outer horizon of these black holes and to check whether they satisfy the chaos bound.
\\
\indent Some interesting phenomena would occur when we pay attention to the near-horizon behavior shown in \cite{Zhao2018StaticEO}. The metric function and potential function of the black hole can be expanded to the second-order on the black hole horizon $r=r_+$:
\bea
\begin{aligned}
f(r)&=f_1(r-r_+)+f_2(r-r_+)^2...
\\
\Phi_e (r)&=\Phi_{e0}+\Phi_{e1}(r-r_+)+\Phi_{e2}(r-r_+)^2...
\end{aligned}\label{efz}
\eea
Then, substitute \meqref{efz} into \meqref{kappa} and \meqref{selambda}
\bea
\begin{aligned}
\kappa& =\frac{1}{2}f_1,
 \\
 \lambda^2&=\left.\frac{12f_1f_2(r-r_+)^2\Phi_{e2}+8f_2^2(r-r_+)^3\Phi_{e2}+f_1^2(\Phi_{e1}+6(r-r_+)\Phi_{e2})}{4(\Phi_{e1}+2(r-r_+)\Phi_{e2})}\right|_{r=r_0}.
\end{aligned}\label{nhlambda}
\eea
Expanding $\lambda^2$ in \meqref{nhlambda} at the horizon $r=r_+$, we can obtain
\bea
\lambda^2=\frac{f_1^2}{4}+\frac{f_1^2\Phi_{e2}}{\Phi_{e1}}(r_0-r_+)+\mathcal{O}((r_0-r_+)^2)\label{nhelambda1},
\eea
which can be rewritten as \cite{Zhao2018StaticEO}
\bea
 \lambda^2=\kappa^2 +\gamma(r_0-r_+)+\mathcal{O}((r_0-r_+)^2),
 \qquad
 \gamma =4\kappa^2\frac{\Phi_{e2}}{\Phi_{e1}}\label{nhelambda2}.
\eea
From \meqref{nhelambda2}, it can be seen that when $\gamma >0$, the chaos bound $\lambda \leq \kappa$ could be violated.
\\
\indent The main reason we consider this dyonic black hole here is that it can provide some special conditions. Since the calculation is very complicated, here we only show the main results of the calculations in the tabular form as shown in Table \ref{tab1}, \ref{tab2}.
\subsubsection*{$\bullet$ Case 1}
We consider the black hole solution given in \meqref{case1} and substitute it into \meqref{kappa} and \meqref{selambda} to calculate the surface gravity of the particles at the black hole horizon and the Lyapunov exponent $\lambda$. Then, the parameter $\gamma$ was calculated through \meqref{nhelambda2}. The calculation results are shown in Table \ref{tab1}.
\\
\begin{table}[h]
\begin{center}
\begin{tabular}{|c|c|c|c|c|c|}
\hline
Black hole & M& Horizon &$\lambda^2$ &$\kappa^2$ &$\gamma$\\
\hline
Case 1 & 1.996&$r_-=0.28617 \quad r_+=3.00002$&0.01283&0.01283&-0.03156\\
\hline
\end{tabular}.
\caption{The results $\lambda^2$, $\kappa^2$ and $\gamma$ of \textbf{Case 1} at the outer horizon $r_+$. As shown, $\lambda = \kappa$ and $\gamma < 0$ mean the chaos bound is satisfied.}
\label{tab1}
\end{center}
\end{table}
\subsubsection*{$\bullet$ Case 2}
Similarly, we also calculate the black hole solution given in \meqref{case2}, and show the results in Table \ref{tab2}.
\\
\\
\begin{table}[h]
\begin{center}
\begin{tabular}{|c|c|c|c|c|c|}
\hline
Black hole & M&Horizon&$\lambda^2$ &$\kappa^2$ &$\gamma$\\	
\hline 	
\textbf{Case 2-1}&6.7&$r_-=0.66823,\quad  r_+=1.54979$&0.01469&0.01469&-0.01537\\
\hline
 \textbf{Case 2-2}&6.8&$r_-=0.53736,\quad r_+=2.07702$&0.00720&0.00720&0.00846\\
 & &$r_1=5.16615,\quad r_2=6.79477$& & & \\
\hline
\textbf{Case 2-3}&7.0&$r_-=0.40827,\quad  r_+=8.34842$&0.00049&0.00049&-0.00039
\\
\hline
\end{tabular}.
\caption{The results $\lambda^2$, $\kappa^2$ and $\gamma$ of three black hole solutions in \textbf{Case 2} at their outer horizon $r_+$. There is always $\lambda = \kappa$ at the outer horizon $r_+$ for ever case. But as we see, $\gamma$ is positive for the black hole in \textbf{Case 2-2}, which means that there may be a violation of the chaos bound $\lambda \leq \kappa$.}
\label{tab2}
\end{center}
\end{table}

\indent The above are the results of all our calculations about $\lambda$, $\kappa$ and $\gamma$ for these dyonic black holes in \meqref{case1} and \meqref{case2}. These calculations show that $\lambda =\kappa$ is universally established on the black hole outside horizon, which means that the chaos bound of $\lambda \leq \kappa$ is satisfied.
\\
\indent However, when considering the near-horizon behavior of the black hole and expanding the correlation function to the second order, we find the case of $\gamma>0$, which indicates that there is a violation of the chaos bound. In order to examine this situation, we need to do more general study about the second-order expansion of these black holes.

\subsection{More discussion on the expansion}
\begin{figure}[htp]
\centering
\subfigure[\scriptsize{\textbf{Case 1}}]{
\includegraphics[scale=0.6]{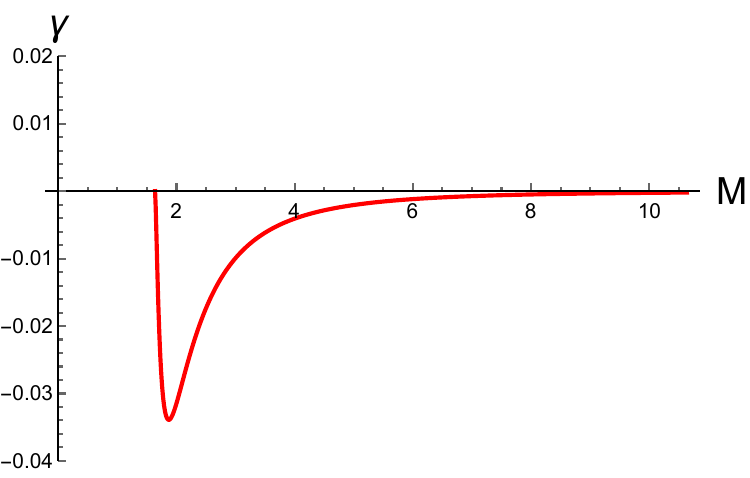}\label{f1a}
}
\quad
\subfigure[\scriptsize{\textbf{Case 2}}]{
\includegraphics[scale=0.6]{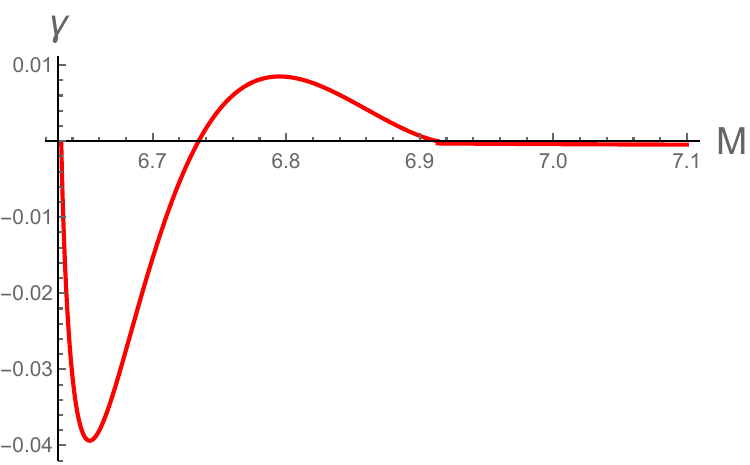}\label{f1b}
}
\caption{The parameter $\gamma$ as a function of the black hole mass $M$. (a) shows no causality violation for the black holes in \textbf{Case 1}. In (b), $\gamma$ can be positive in the range $M \in (6.73442,6.9135)$, which implies that the chaos bound $\lambda \leq \kappa$ is violated in \textbf{Case 2}.}\label{gp}
\end{figure}

To more clearly show the anomalous behavior of $\gamma$ in the second-order expansion, we will discuss these two sets of metric functions and their corresponding potential functions above (\meqref{case1} and \meqref{case2}), and then plot $\gamma$ as a function of $M$ in Figure \ref{gp}.
\\
\indent There is always $\gamma <0$ in \mpref{f1a}, which means the chaos bound $\lambda \leq \kappa$ is not violated for \textbf{Case 1}. However, as shown in \mpref{f1b}, for the black holes with $M \in (6.73442,6.9135)$ in \textbf{Case 2}, the chaos bound can be violated in the near horizon region.

\subsection{Numerical analysis}\label{toy model}
To study the effect of the violation of chaos bound more, we consider a toy model that describes the near-horizon geometry with external potentials and study the Poincar\'e section of the particle motion in this model. The Lagrangian of particles can be written as
\be
\mathcal{L}=-\sqrt{f(x)-\frac{\dot{x}^2}{f(x)}-\dot{y}^2}-\omega(A(x)+B(y)),\label{tmL}
\ee
where ``$\cdot$'' denotes derivative with respect to the coordinate time $t$, the metric function $f(x)=2\kappa (x-x_h)$, $\kappa$ is the surface gravity, $x_h$ is the radius of black hole horizon. (Note that the near-horizon expansion of the metric \meqref{Dmf} yields the form $f(x)=2\kappa (x-x_h)$.) $A(x)$ and $B(y)$ are external potentials, $\omega$ is the coupling coefficient of particle with external potential\footnote{The relation between $\omega$ and the external potentials($A(x)$ and $B(y)$) can be regarded as the relation between the charge of charged particle for the electrial potential. }.
\\
\indent From the Lagrangian \meqref{tmL}, we can derive the generalized momentum
\be
\begin{aligned}
P_x&=\frac{\dot{x}}{f(x)\sqrt{f(x)-\frac{\dot{x}^2}{f(x)}-\dot{y}^2}},
\\
P_y&=\frac{\dot{y}}{\sqrt{f(x)-\frac{\dot{x}^2}{f(x)}-\dot{y}^2}}.
\end{aligned}
\ee
The Hamiltonian of particle (that is, the energy $E$) is
\be
\begin{aligned}
	E&=P_x\dot{x}+P_y\dot{y}-\mathcal{L}
	\\
	&=\sqrt{f(x)(P_x^2f(x)+P_y^2+1)}+\omega (A(x)+B(y)).
\end{aligned}\label{nenergy}
\ee
Correspondingly, the equation of motion can be written as
\be
\begin{aligned}
	\frac{dx}{dt}&=\frac{\partial{E}}{\partial{P_x}}=P_xf(x)\sqrt{\frac{f(x)}{P_x^2f(x)+P_y^2+1}}
	\\
	\frac{dP_x}{dt}&=-\frac{\partial{E}}{\partial{x}}=-P_x^2\partial_xf(x)\sqrt{\frac{f(x)}{P_x^2f(x)+P_y^2+1}}-\frac{P_y^2\partial_xf(x)}{2f(x)}\sqrt{\frac{f(x)}{P_x^2f(x)+P_y^2+1}}
	\\
	&-\frac{\partial_xf(x)}{2f(x)}\sqrt{\frac{f(x)}{P_x^2f(x)+P_y^2+1}}-\omega \partial_xA(x)
	\\
	\frac{dy}{dt}&=\frac{\partial{E}}{\partial{P_y}}=P_y\sqrt{\frac{f(x)}{P_x^2f(x)+P_y^2+1}}
	\\
	\frac{dP_y}{dt}&=-\frac{\partial{E}}{\partial{y}}=-\omega \partial_yB(y)
\end{aligned}\label{neom}
\ee
\indent Similarly as in \cite{Hashimoto2017UniversalityIC,Dalui2018PresenceOH}, we consider the external potential similar to the harmonic oscillator potential
\be
\begin{aligned}
	A(x)&=a(x-x_c)^2+b(x-x_c)^4,
	\\
	B(y)&=y^2,
\end{aligned}
\ee
where $x_c$ is the center position of $A(x)$. We set $x_h=0$ and $x_c=1$, then two cases that the chaos bound is violated or not can be decided by adjusting the parameters $a$ and $b$. To avoid the particle falling into horizon, the particle energy $E$ should have an upper bound $E_{max}$ (See Appendix \ref{toy model Parameters} for the detailed derivation of $a$, $b$ and $E_{max}$). Here, we consider the values of $a$ and $b$ as following:
\\
\\
\indent \emph{1. Chaos bound can be violated }
\begin{equation*}
a=1.3, \qquad b=-0.3.
\end{equation*}
\\
\indent \emph{2. Chaos bound not violated }
\begin{equation*}
a=1.1, \qquad b=-0.1.
\end{equation*}
The other parameters are chosen as $\kappa=\frac{1}{2}$ and $\omega=E_{max}=30$.
\\
\indent The equation of motion \meqref{neom} can be solved numerically. About the initial condition, we set $y(0)=0$, $P_x(0)=0$, and the same $x(0)$ for two cases, and the corresponding $P_y(0)$ can be obtained by \meqref{nenergy}. Then we can define the Poincar\'e section in $(x,P_x)$ plane with $y=0$ and $P_y<0$. In the following figures, we put the Poincar\'e section of the system (a=1.3, b=-0.3) where the chaos bound can be violated on the left, and the non-violation system (a=1.1, b=-0.1) on the right.
\begin{figure}[htp]
	\centering
	\subfigure[\scriptsize{a=1.3, b=-0.3}]{
	\includegraphics[scale=0.34]{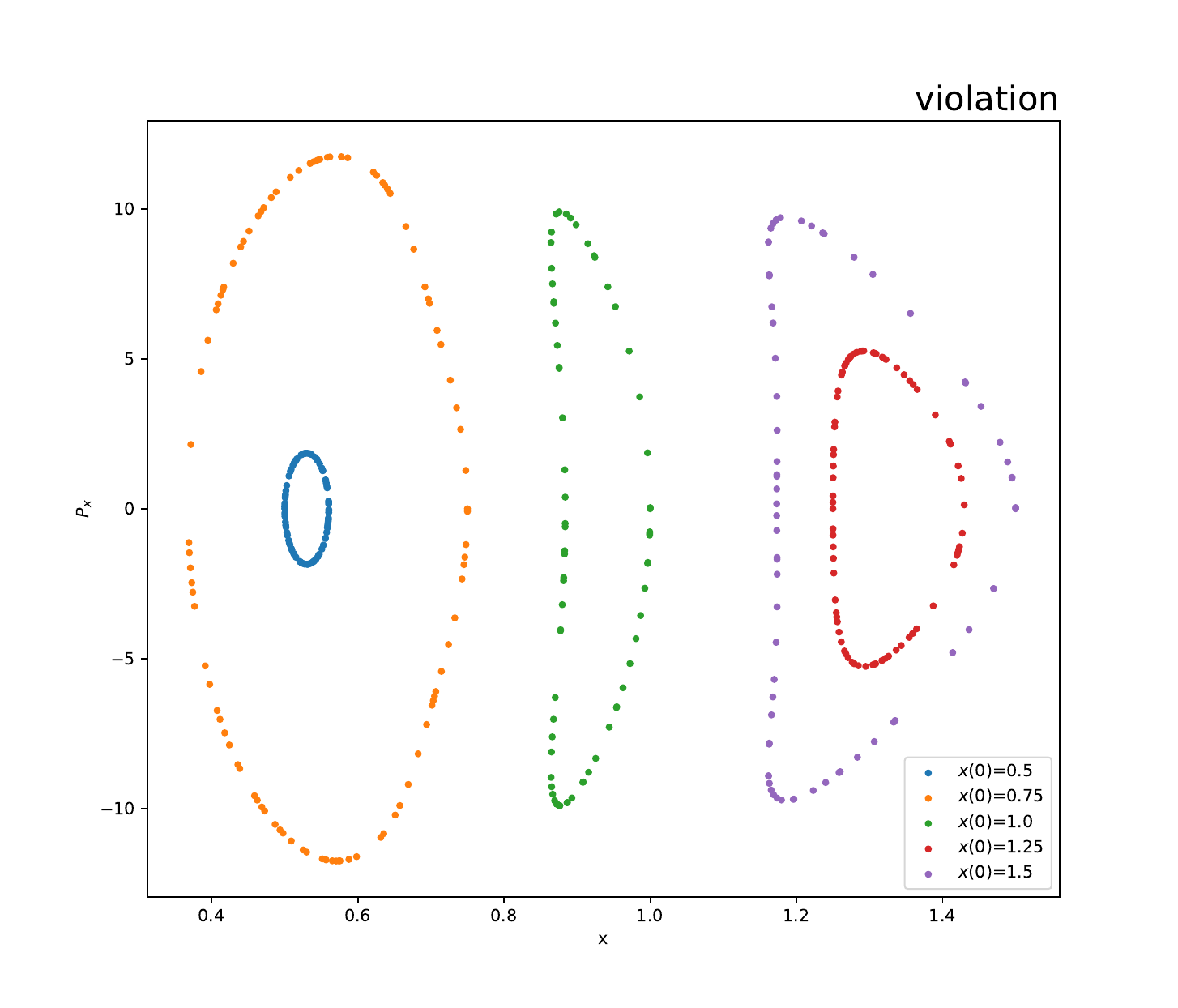}
	}
	\quad
	\subfigure[\scriptsize{a=1.1, b=-0.1}]{
	\includegraphics[scale=0.34]{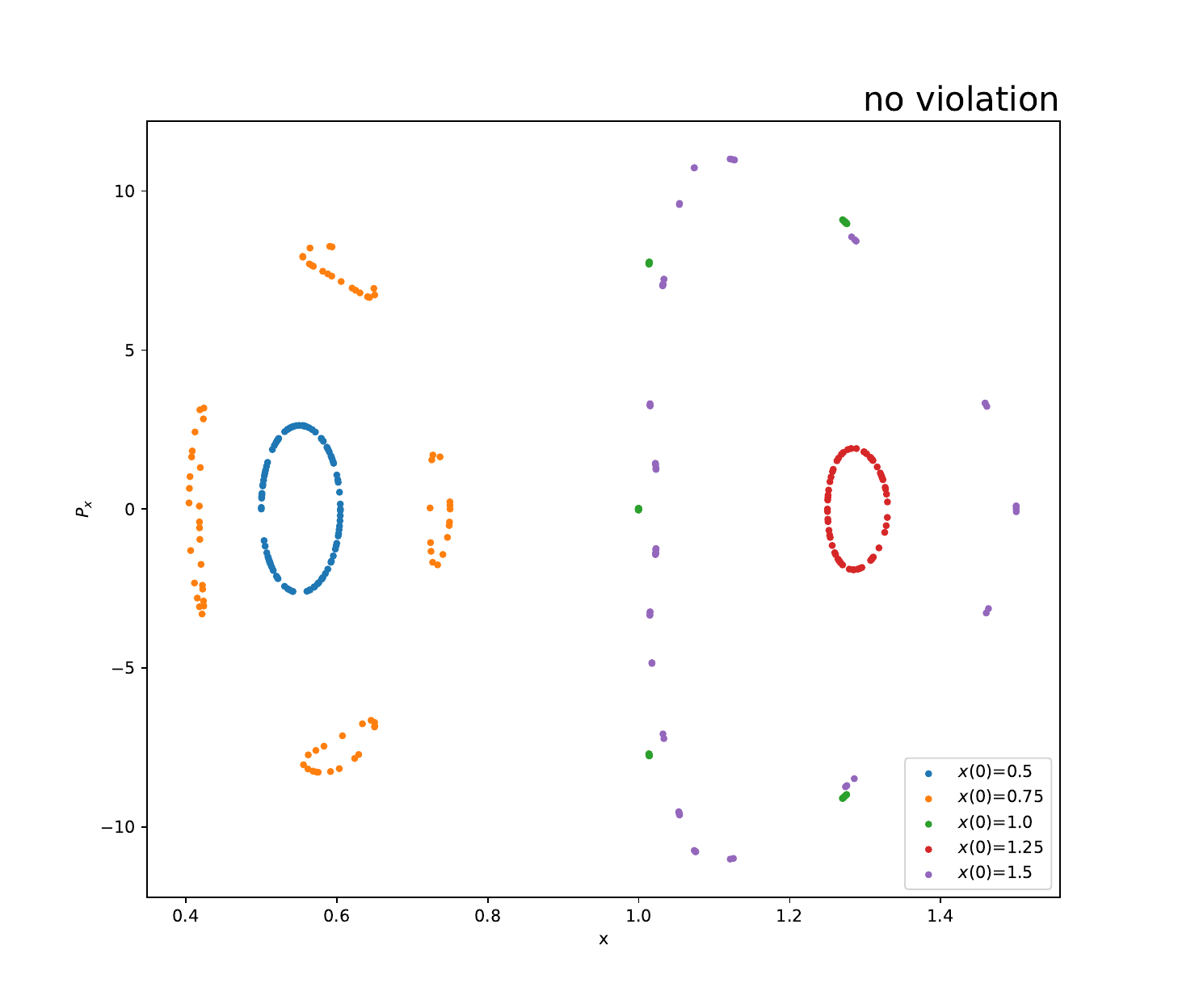}
	}
	\caption{E=18: The Poincar\'e section in $(x,P_x)$ plane with $y=0$ and $P_y<0$. In the left figure, there are many closed rings which means these orbits are quasi-periodic. Many closed curves can be found in the right figure too, but they are more irregular.}\label{E18ps}
\end{figure}
\begin{figure}[htp]
	\centering
	\subfigure[\scriptsize{a=1.3, b=-0.3}]{
	\includegraphics[scale=0.34]{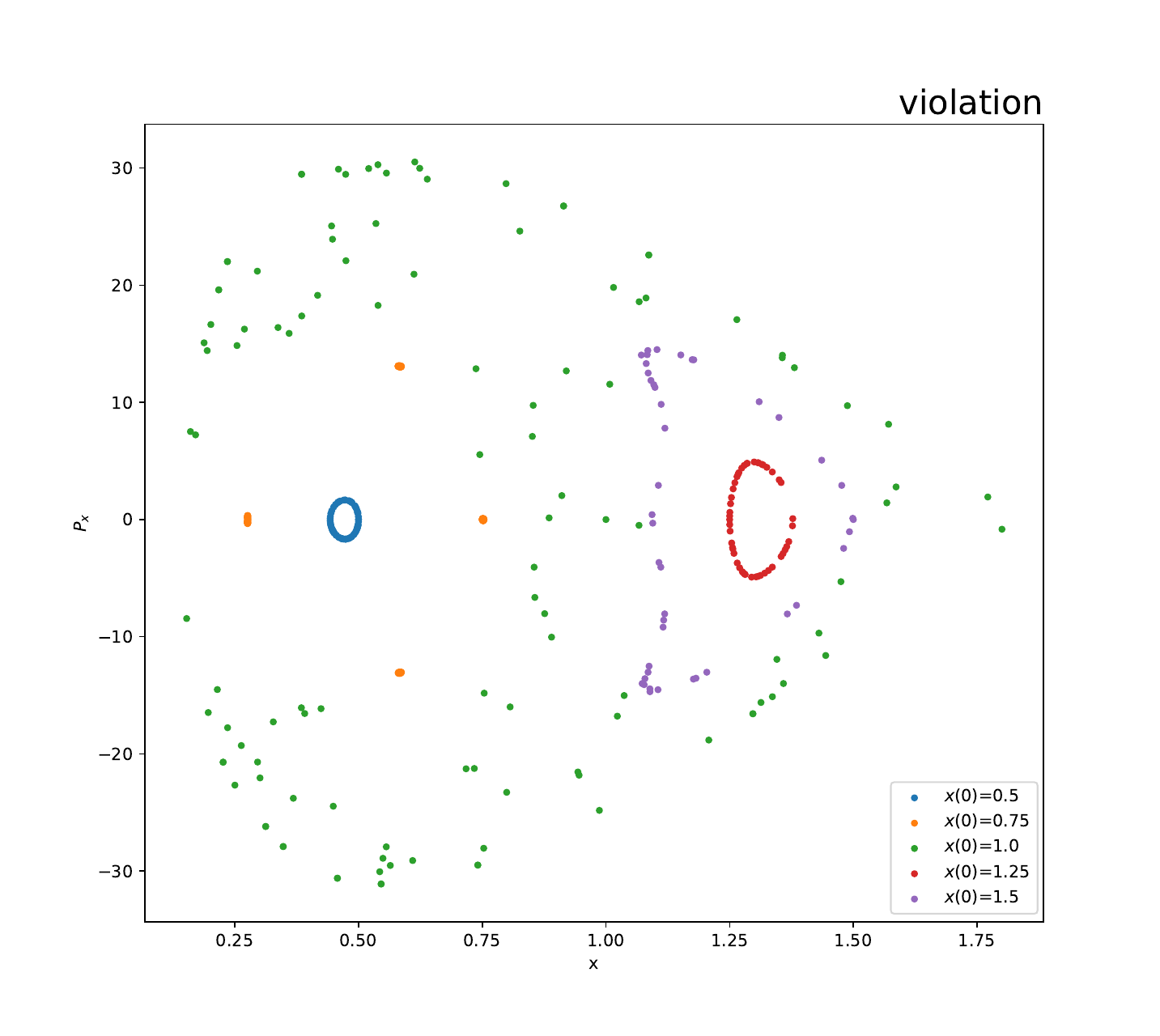}
	}
	\quad
	\subfigure[\scriptsize{a=1.1, b=-0.1}]{
	\includegraphics[scale=0.34]{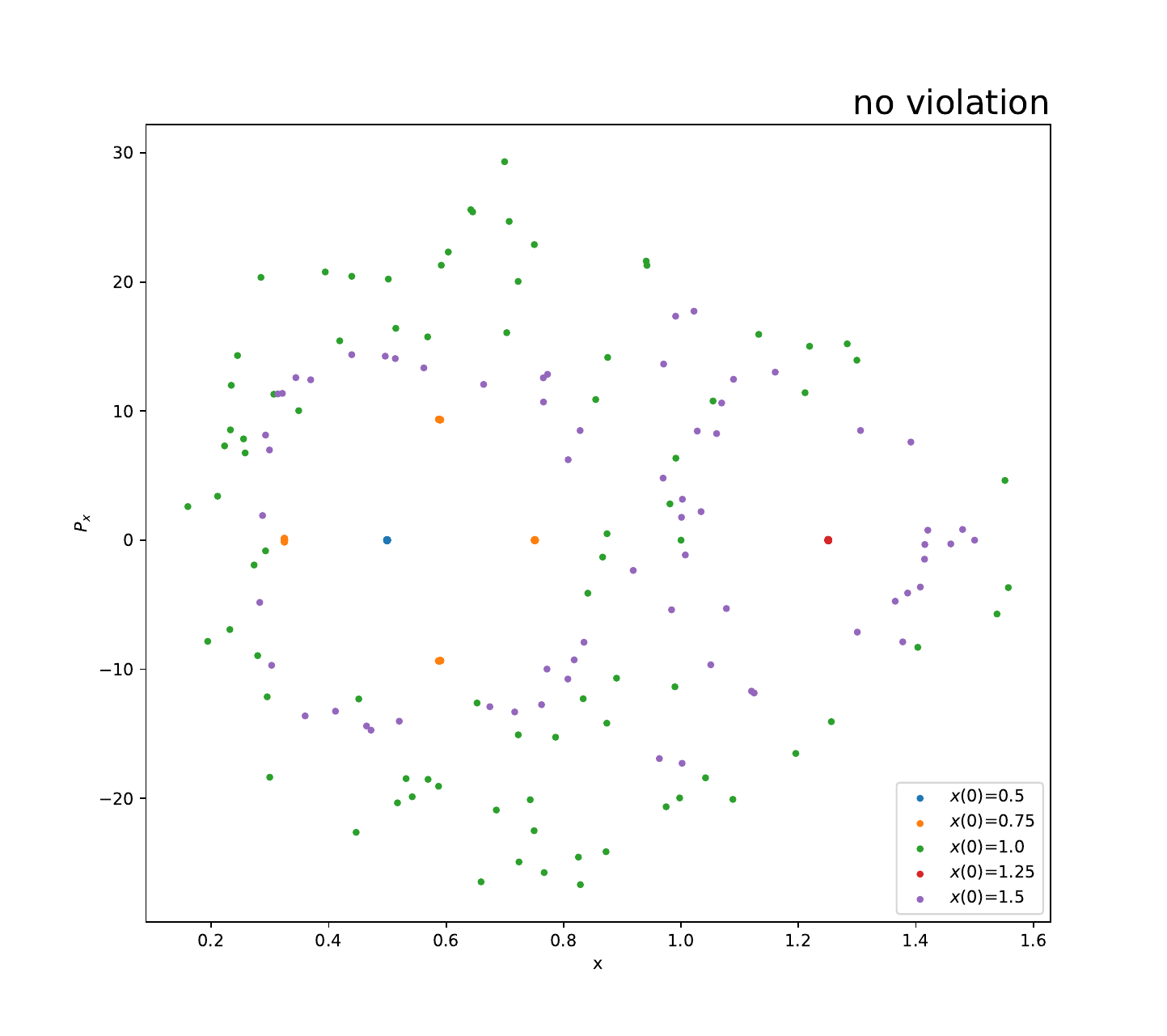}
	}
	\caption{E=25: The Poincar\'e section in $(x,P_x)$ plane with $y=0$ and $P_y<0$. In both two figures, the points of $x(0)=1.0$ are dispersed which means the orbits are chaotic. When $x(0)=0.5$ and $x(0)=1.25$, the points form rings in the left figure, and thay are single points on the right. In the left figure, the points of $x(0)=1.5$ form a closed curve. There is a chaotic trajectory of $x(0)=1.5$ in the right figure.}\label{E25ps}
\end{figure}
\\
\indent In \mpref{E18ps}, we set $E=18$ and show the Poincar\'e section. The particle motion seems to be more closer to chaos in the system where the chaos bound cannot be violated since the section in Figure 2(b) is more irregular than in Figure 2(a). We should point out that the reason of this difference maybe the system($a=1.3$, $b=-0.3$) has a bigger potential $A(x)$ than another. For the same $E$, the particle will have smaller kinetic energy which may weaken the chaos in particle motion. As for why the chaos is not strengthened in the system where the chaos bound can be violated, we speculate that the low energy results in it. The greater the energy $E$, the closer the particle can approach the horizon until it falls into the black hole. The violation of chaos bound we studied is a near-horizon behavior, so its reinforcement of chaos will work when the particles are close to horizon. To verify this, we examined the Poincar\'e section with higher energy. The situation of $E=25$ is shown in \mpref{E25ps}. In this figure, we can see the chaos is strengthened in the system with $a=1.3$ and $b=-0.3$. The points of $x(0)=0.5$ and $x(0)=1.25$ form closed curves in Figure 3(a) which indicates quasi-periodic orbits, and in Figure 3(b) they are single points which represent periodic orbits. As we expected, the chaos in system ($a=1.3$, $b=-0.3$) has been strengthened, because of the bigger $E$ results in particle motion approaching horizon. Meanwhile, there is an opposite result of points with $x(0)=1.5$, as the system ($a=1.3$, $b=-0.3$) with stronger $A(x)$. These points in Figure 3(a) form a closed curve, and they are dispersed which means it is a chaotic orbit in the system ($a=1.1$, $b=-0.1$).
\\
\indent To further examine our intuition, we set E=29.9 in \mpref{E29ps}, which will make the particles move close enough to the horizon to make the enhancement of chaos in system ($a=1.3, b=-0.3$) more obvious. At the same time, we calculated more data. As shown in \mpref{E29ps}, there are many chaotic orbits in Figure 4(a), and the points with same values of $x(0)$ in Figure 4(b) are regular which means these orbits are not chaotic. The chaos in the system ($a=1.3$, $b=-0.3$) has been significantly strengthened. Although the stronger potential $A(x)$ in the system ($a=1.3$, $b=-0.3$) will weaken the chaos, we still see that the chaos is strengthened in \mpref{E29ps}, which undoubtedly shows in a system where the chaos bound can be violated, chaos can indeed be strengthened.

\begin{figure}[htp]
	\centering
	\subfigure[\scriptsize{a=1.3, b=-0.3}]{
	\includegraphics[scale=0.34]{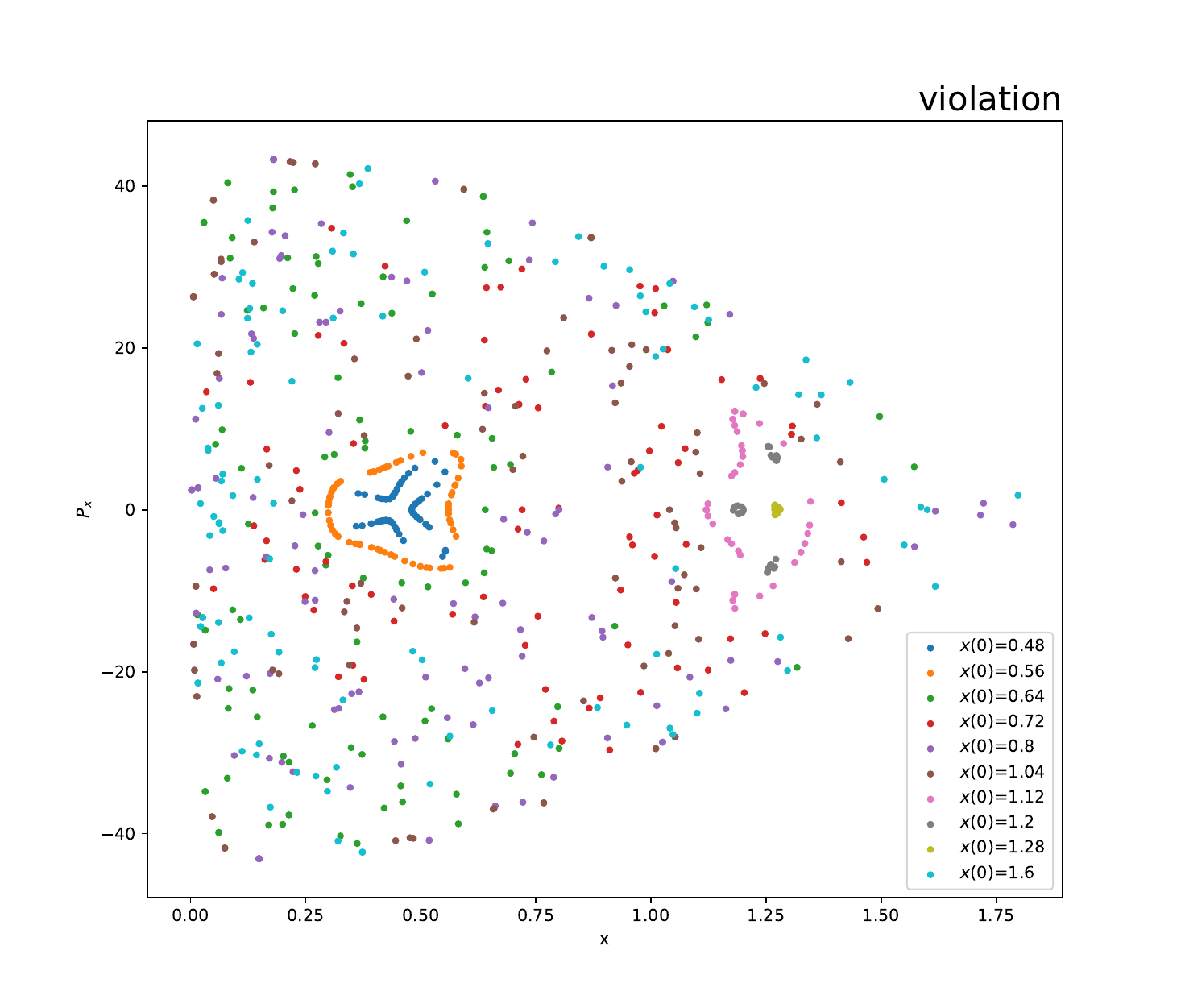}
	}
	\quad
	\subfigure[\scriptsize{a=1.1, b=-0.1}]{
	\includegraphics[scale=0.34]{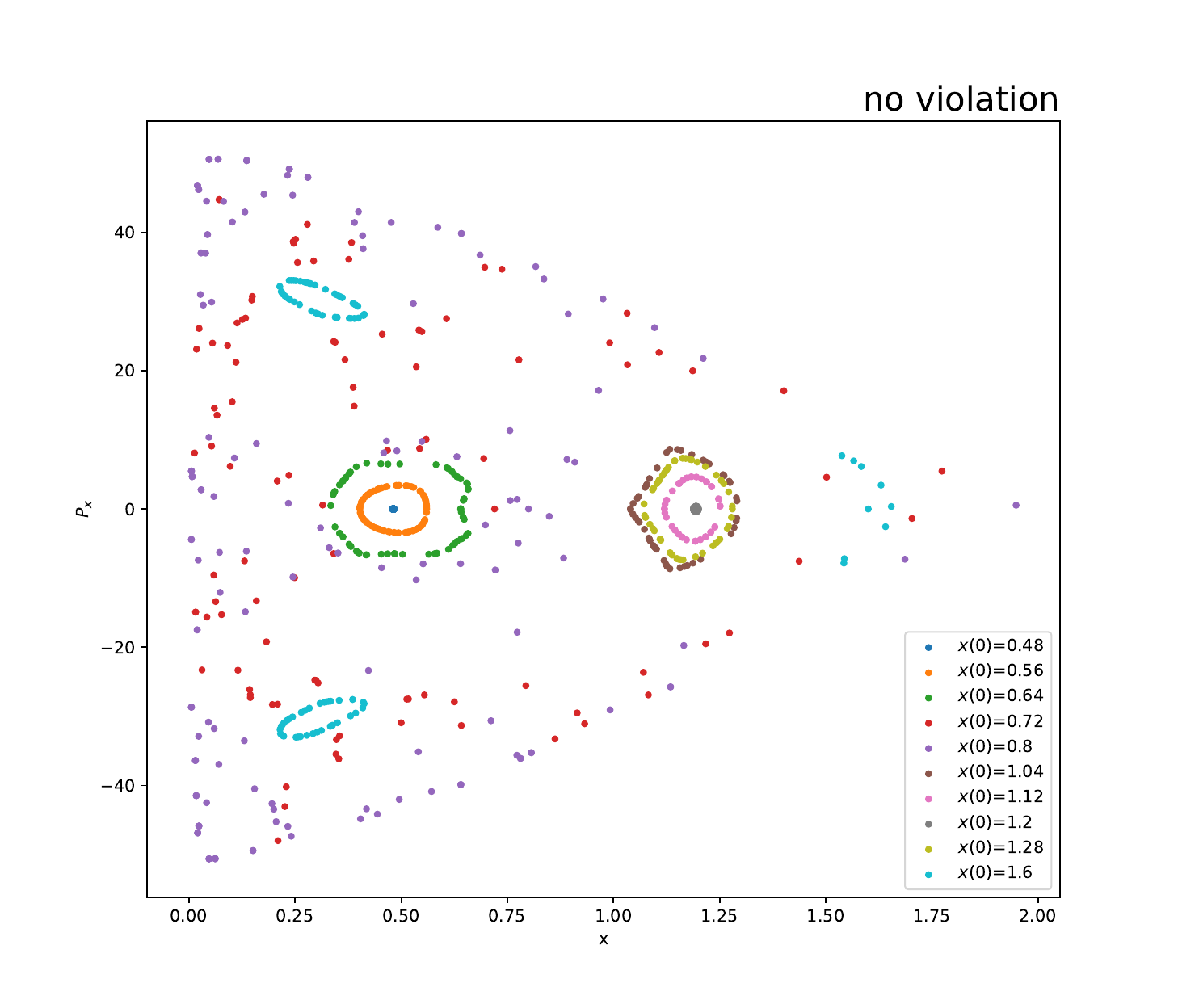}
	}
	\caption{E=29.99: The Poincar\'e section in $(x,P_x)$ plane with $y=0$ and $P_y<0$. In the left figure, the points of $x(0)=(0.64,\ 0.72,\ 0.8,\ 1.04,\ 1.6)$ are dispersed, and the other points form closed curves. At right, there are two single points of $x(0)=0.48$ and $x(0)=1.2$. The points of $x(0)=(0.56,\ 0.64,\ 1.04,\ 1.12,\ 1.28,\ 1.6)$ form closed curves, and the other points of $x(0)=(0.72,\ 0.8)$ are dispersed.}\label{E29ps}
\end{figure}

\section{Geodesic motion of neutral particles}\label{Geodesic motion of neutral particles}
In this section, we study the Lyapunov exponent of the geodesic motion of neutral particles near the black hole, including circular geodesic motion and radial falling. We extend the static equilibrium of charged particles in Section \ref{Static equilibrium of charged particles} beyond the horizon, and take the ``maximal'' Lyapunov exponent as $\lambda_s$ which can be obtained from the static equilibrium. From \meqref {selambda}, we obtain the formula of $\lambda_s$ for charged black holes
\bea
\lambda_s=\sqrt{f\left(r \right)\sqrt{f\left(r \right)}\left(\frac{\Phi_e^{''}\left(r \right)}{\Phi_e^{'}\left(r \right)}\left(\sqrt{f\left(r \right)} \right)^{'}-\left(\sqrt{f\left(r \right)} \right)^{''} \right)}.\label{ls}
\eea
\indent Arbitrary location outside the horizon where the effective potential $V_{eff}$ has a maximum, the Lyapunov exponent $\lambda_s$ obtained from static equilibrium is at its maximal value. When we consider $\lambda_s$ at the horizon, the value of $\lambda_s$ will return to the surface gravity $\kappa$. We propose a hypothesis here: the ``maximal'' Lyapunov exponent $\lambda_s$ is an inherent property determined by the nature of the black hole\footnote {This is our assumption that \meqref{ls} is maximal at any position outside the black hole. In other words, we assume there is a renormalization group flow of $\lambda_s$ outside the black hole. It can flow to $\lambda_s = \kappa$ at the event horizon. We will compare \meqref{ls} with the Lyapunov exponents obtained from particle circular geodesic motion \cite{cardoso2008geodesic,pradhan2012stability,pradhan2014circular} and radial falling.}. It is worthwhile to explore the connection between the ``maximal'' Lyapunov exponent $\lambda_s$ with the Lyapunov exponent of particle geodesic motion.
\subsection{Circular geodesic motion of neutral particles}
A 4-dimensional spherically symmetric black hole is considered here. Its metric is
\bea
ds^2=-f\left(r\right)dt^2+\frac{dr^2}{f\left(r\right)}+d\Omega^2_2\label{4dbh}.
\eea
When we focus on a neutral test particle moving on the equatorial plane of this black hole ($\theta =\frac{\pi}{2}, \dot{\theta}=0$), its Lagrangian can be written as
\bea
\mathcal{L}=\frac{1}{2}\left[g_{tt}\dot{t}^2+g_{rr}\dot{r}^2+g_{\phi \phi}\dot{\phi}^2 \right],
\eea
where the dot `` . '' denotes derivative with respect to the proper time `` $\tau$ ''.
\\
\indent According to the generalized momentum expression $p_q=\frac{\partial L}{\partial \dot{q}}$, we can obtain its generalized momentum as
\bea
\begin{aligned}
&p_t=-f\left(r\right)\dot{t}=-E=Const,
\\
&p_{\phi}=r^2\dot{\phi}=L=Const,
\\
&p_r=\frac{\dot{r}}{f\left(r\right)},
\end{aligned}\label{cgm}
\eea
where $E$ is the particle's energy and $L$ is the angular momentum of the particle. Using the normalization with four-velocity $g_{\mu \nu}u^{\mu}u^{\nu}=\eta$
\bea
-E\dot{t}+L\dot{\phi}+\frac{\dot{r}^2}{f\left(r\right)}=\eta,
\eea
we can obtain
\bea
\dot{r}^2=E^2-\left(\frac{L^2}{r^2}-\eta \right)f\left( r\right)\label{rd}.
\eea
Note that $\eta=+1,-1,0$ corresponds to space-like, time-like and null geodesics respectively.
\\
\indent For the circular geodesic motion of neutral particles, we have the Euler-Lagrange equation
\bea
\frac{dp_q}{d\tau}=\frac{\partial \mathcal{L}}{\partial q}\label{elp}.
\eea
Setting the phase space variables $X_i\left(t\right)=\left(p_r,r\right)$ and considering the particles moving in a circular orbit with a radius of $r=r_0$, we have two equations
\bea
\frac{dp_r}{d\tau}=\frac{\partial{\mathcal{L}}}{\partial r} \quad \textrm{and} \quad \frac{dr}{d\tau}=\frac{p_r}{g_{rr}}.
\eea
Then, the Jacobian matrix of particle motion can be obtained from \meqref{jmd}
\bea
K_{ij}=\left.\left|
\begin{array}{cc}
0&\frac{d}{dr}\left(\frac{\partial \mathcal{L}}{\partial r}\right)\\
\frac{1}{g_{rr}}&0\\
\end{array}
\right|\right|_{r=r_0}.
\eea
The eigenvalues of this matrix can give the expression of the proper time Lyapunov exponent of circular geodesic motion $\lambda_p$, and we observe that $\lambda_p$ satisfies \cite{pradhan2012stability}
\bea
\lambda_p^2=\left.\frac{1}{g_{rr}}\frac{d}{dr}\left(\frac{\partial \mathcal{L}}{\partial r}\right)\right|_{r=r_0}\label{cgp1}.
\eea
From the Lagrange's equation of geodesic motion
\bea
\frac{d}{d\tau}\left(\frac{\partial \mathcal{L}}{\partial{\dot{r}}}\right)-\frac{\partial \mathcal{L}}{\partial{r}}=0,
\eea
and the formula
\bea
\frac{d}{d\tau}\left(\frac{\partial \mathcal{L}}{\partial{\dot{r}}}\right)=\frac{d}{d\tau}\left(-g_{rr}\dot{r}\right)=-\dot{r}\frac{d}{dr}\left(g_{rr}\dot{r}\right)=-\frac{1}{2g_{rr}}\frac{d}{dr}\left(g_{rr}^2\dot{r}^2\right),
\eea
the expression of $\frac{\partial \mathcal{L}}{\partial{r}}$ can be written as
\bea
\frac{\partial \mathcal{L}}{\partial{r}}=-\frac{1}{2g_{rr}}\frac{d}{dr}\left(g_{rr}^2\dot{r}^2\right).\label{pl}
\eea
 For the circular geodesic motion, we have the circular geodesic condition \cite{cgc}
\bea
\dot{r}^2=\left(\dot{r}^2 \right)^{'}=0.
\eea
Under this condition, we substitute \meqref{pl} into \meqref{cgp1}, then the proper time Lyapunov exponent $\lambda_p$ in \meqref{cgp1} can be reduced to
\bea
\lambda_p=\sqrt{\frac{\left(\dot{r}^2\right)^{''}}{2}}\label{cgp2}.
\eea
If we consider an alternative form of \meqref{elp}
\bea
\frac{dP_q}{dt}=\frac{d\tau}{dt}\frac{\partial \mathcal{L}}{\partial q}\label{elc},
\eea
we can express the Lyapunov exponent $\lambda_c$ in term of coordinate time \cite{cardoso2008geodesic}
\bea
\lambda_c=\sqrt{\frac{\left(\dot{r}^2\right)^{''}}{2\dot{t}^2}}\label{cgc}.
\eea
In \meqref{cgm}, $\dot{t}=\frac{E}{f\left(r\right)}$, so we obtain the relation between \meqref{cgp2} and \meqref{cgc}
\bea
\lambda_c=\frac{f\left(r\right)}{E}\lambda_p.
\eea
Both $f\left(r\right)$ and $E$ are real, so the properties of $\lambda_p$ and $\lambda_c$ are closely related. It is evident that the Lyapunov exponents $\lambda_p$ and $\lambda_c$ are the most important parameters directly verifying the stability of the motion. Only when $\lambda_p$ and $\lambda_c$ are both real, they result in unstable geodesic motion. In contrast, when one of the $\lambda_p$ and $\lambda_c$ is imaginary, the circular geodesic motion is stable; when $\lambda_p=0$ or $\lambda_c=0$, the circular geodesic motion is marginal, which means the circular geodesic motion can be easily broken.
\\
\indent The Lyapunov exponent is a measure of the deviation in the time evolution of two adjacent trajectories in phase space, obviously it depends on the time coordinate. In the following, we will study the coordinate time Lyapunov exponent $\lambda_c$ of the neutral particle's geodesic motion.
\subsubsection{Circular time-like geodesic}
For time-like geodesic with $\eta=-1$, we have
\bea
\dot{r}^2=E^2-\left(\frac{L^2}{r^2}+1 \right)f\left( r\right)\label{tled}.
\eea
The circular geodesic condition $\dot{r}^2=\left(\dot{r}^2 \right)^{'}=0$ yields \cite{cardoso2008geodesic}
\bea
E^2=\frac{2f^2\left(r\right)}{2f\left(r\right)-rf^{'} \left(r\right)}, \quad L^2=\frac{r^3f^{'}\left(r\right)}{2f\left(r\right)-rf^{'}\left(r\right)}.\label{tlEL}
\eea
After substituting the metric functions of the four black holes given by \meqref{case1} and \meqref{case2} into \meqref{tlEL} and \meqref{tled}, the coordinate time Lyapunov exponent $\lambda_c$ of these circular time-like geodesic motion can be given by
\meqref{cgc}.
\\
\indent To have a clear picture, we compare the circular geodesic motion of RN black hole with the black hole solutions given by \meqref{case1} and \meqref{case2}. The metric function and the electrial function of RN black hole are
\bea
\begin{aligned}
f\left(r\right)&=1-\frac{2M}{r}+\frac{Q^2}{r^2},
\\
\Phi_e\left(r\right)&=\frac{2Q}{r},
\end{aligned}\label{RN}
\eea
where $M$ is the mass of black hole, and $Q$ is the charge. For RN black hole, $\lambda_c^2$ and $\lambda_s^2$ can be calculated by \meqref{cgc} and \meqref{ls}
\bea
\begin{aligned}
	\lambda_c^2&=\frac{-Mr^3+6M^2r^2-9MQ^2r+4Q^2}{r^6},
	\\
	\lambda_s^2&=\frac{M^2-Q^2}{r^4}.
\end{aligned}
\eea
\indent We want to explore the relationship between the circular time-like geodesic motion's Lyapunov exponent $\lambda_c$ and ``maximal'' Lyapunov exponent $\lambda_s$ defined from the static equilibrium, so we plot figures of $\lambda_c^2$ and $\lambda_s^2$ in the region where the circular geodesic motion exists.

\begin{figure}[htpb]
\centering
\subfigure[$5.64575<r<16$]{
\includegraphics[scale=0.25]{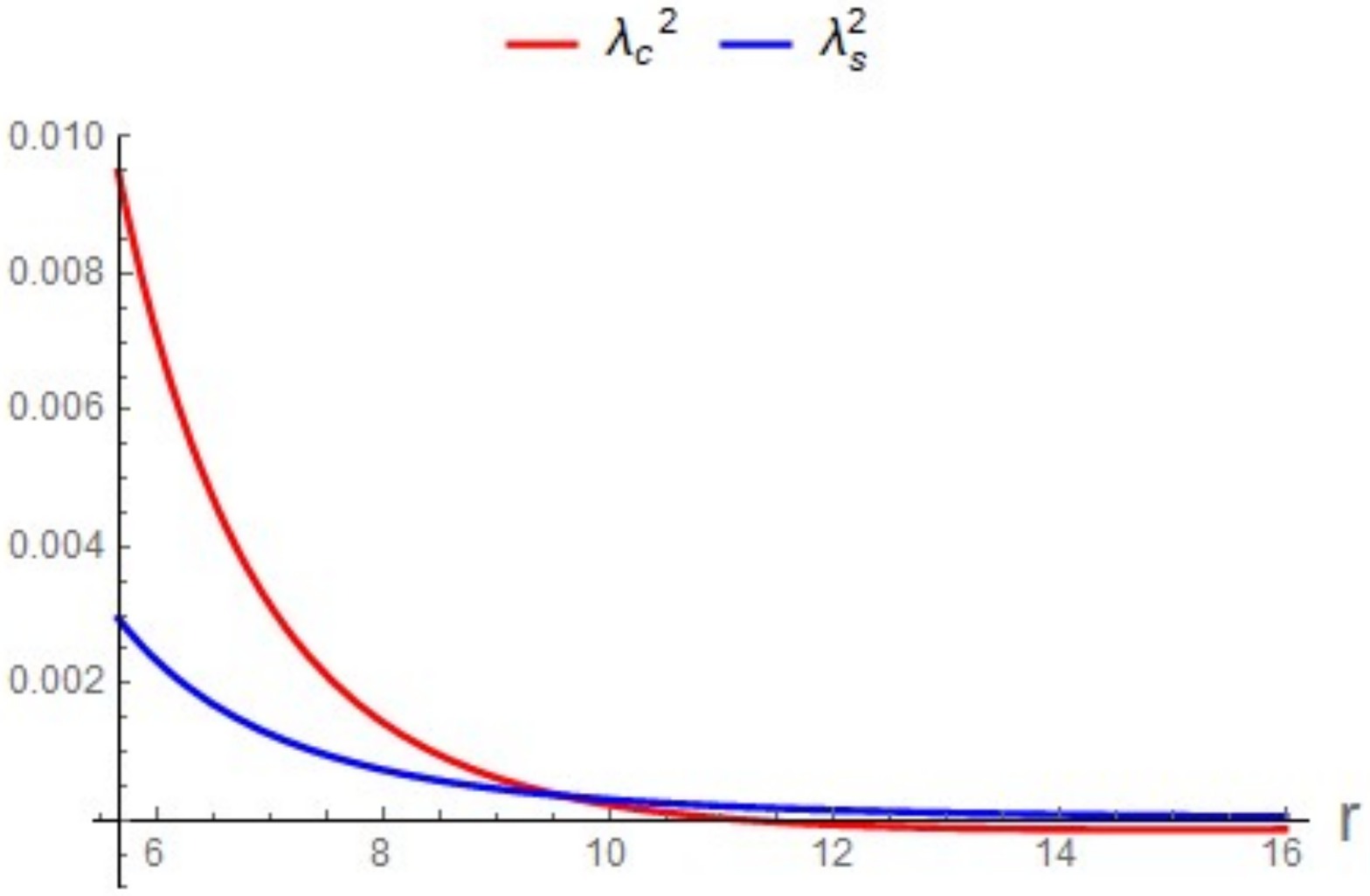}
}
\quad
\subfigure[$10<r<36$]{
\includegraphics[scale=0.25]{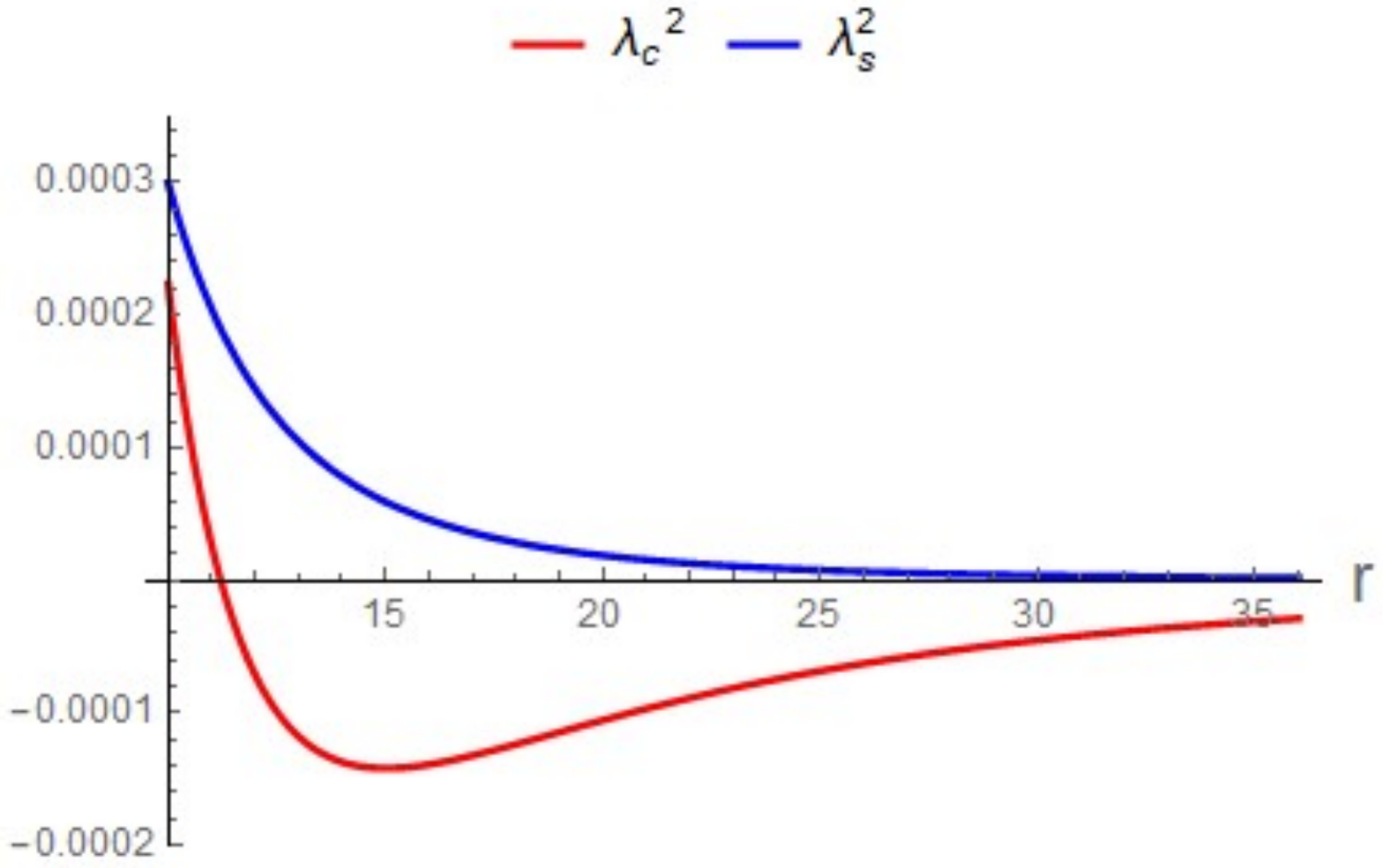}
}
\caption{$\lambda_c$ and $\lambda_s$ are defined by \meqref{cgc} and \meqref{ls} as functions of radial coordinate $r$. For the RN black hole with $M=2$ and $Q=1$, $\lambda_c^2$ decreases from the radius of the innermost circular orbit (the beginning of the coordinate axis). $\lambda_s^2$ decreases too, but $\lambda_s^2$ is always positive in contrast to $\lambda_c^2$. Near the innermost orbital radius, we have $\lambda_c^2 > \lambda_s^2$. Away from the black hole, $\lambda_c^2$ is negative, which means the circular geodesic motion is stable.}\label{cRN}
\end{figure}
\begin{figure}[htpb]
\centering
\includegraphics[scale=0.6]{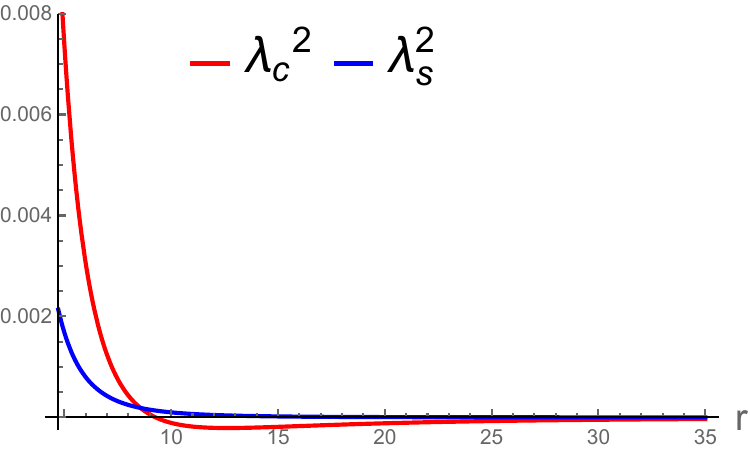}
\caption{For the \textbf{Case 1}, the black hole defined by \meqref{case1} with $M=1.996$, $\lambda_c^2$ and $\lambda_s^2$ are similar in the RN black hole.}\label{cCase1}
\end{figure}
\begin{figure}[htpb]
\centering
\subfigure[\scriptsize{The range $2.19387<r<2.84801$.}]{
\includegraphics[scale=0.36]{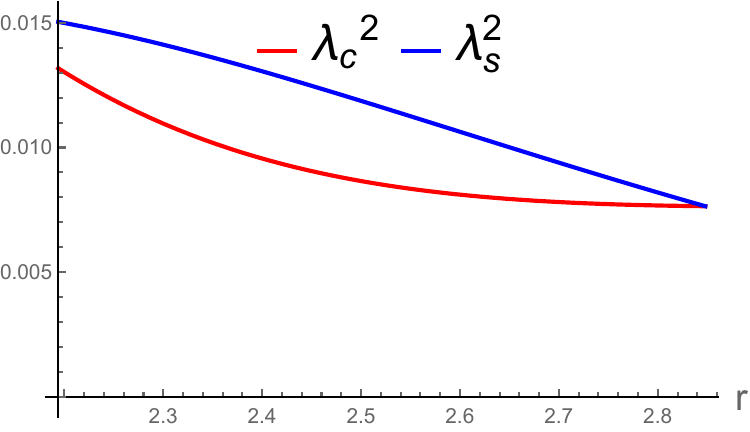}\label{c21a}
}
\quad
\subfigure[\scriptsize{The range $6.14209<r<6.44301$. $\lambda_s$ can't be defined in this range.}]{
\includegraphics[scale=0.36]{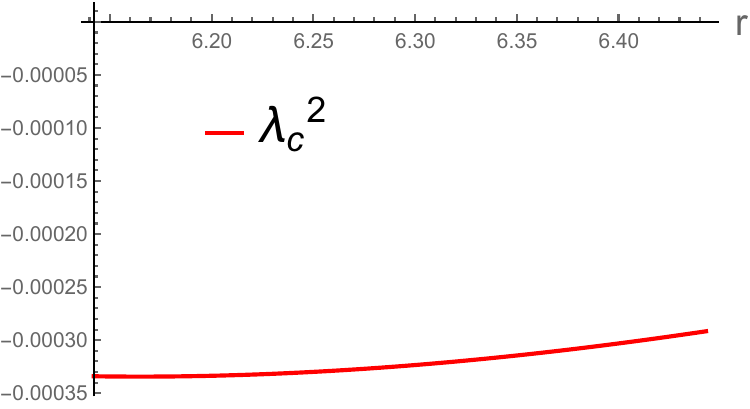}\label{c21b}
}
\quad
\subfigure[\scriptsize{The range $r>12.4354$. $\lambda_s$ can't be defined in this range.}]{
\includegraphics[scale=0.36]{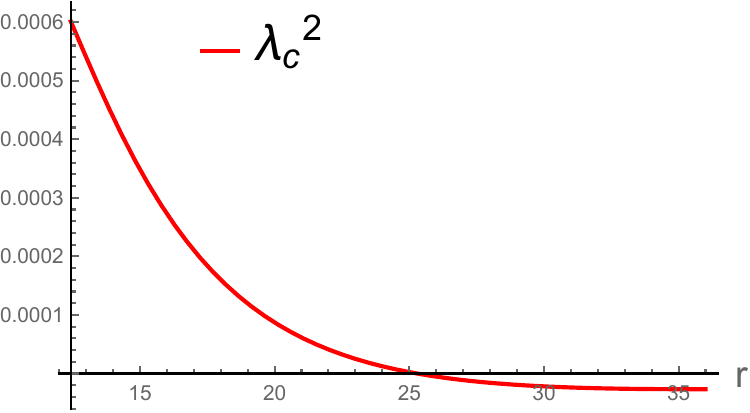}\label{c21c}
}
\caption{For the \textbf{Case 2-1}, the black hole defined by \meqref{case2} with $M=6.7$, the circular time-like geodesic motion exists in three discontinuous range. As shown in (b), there is $\lambda_c^2 <0$ in the range $r \in (6.14209,6.44301)$ near the horizon, which means that in this range there can be a stable time-like circular geodesic motion. In (c), $\lambda_c^2$ has a RN-like behavior.}\label{cCase2-1}
\end{figure}
\begin{figure}[htpb]
\centering
\includegraphics[scale=0.6]{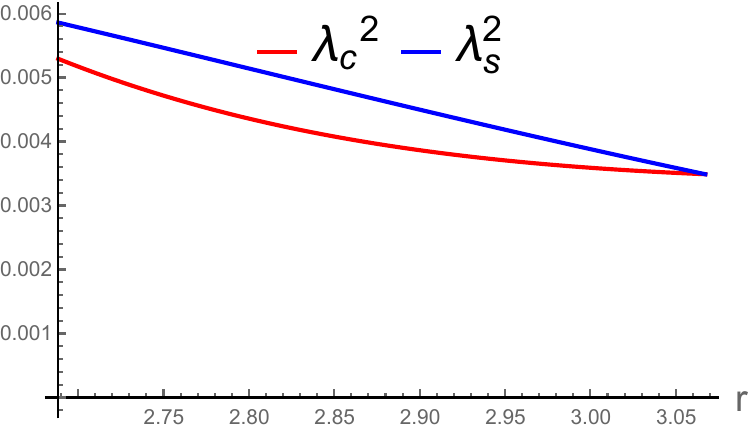}
\caption{For the \textbf{Case 2-2}, the black hole defined by \meqref{case2} with $M=6.8$, there is always $\lambda_s^2>\lambda_c^2$ in the range $2.68843<r<3.06723$.}\label{cCase2-2}
\end{figure}
\begin{figure}[htpb]
\centering
\includegraphics[scale=0.6]{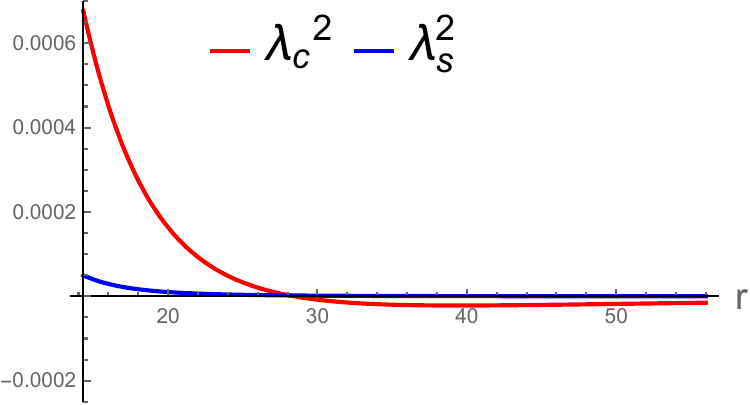}
\caption{For the \textbf{Case 2-3}, the black hole defined by \meqref{case2} with $M=7$, the behavior of $\lambda_c^2$ and $\lambda_s^2$ is similar to that of RN black hole.}\label{cCase2-3}
\end{figure}

\indent Firstly, we focus on the stability of circular motion which can be expressed by $\lambda_c^2$. For the RN black hole, as we show in \mpref{cRN}, $\lambda_c^2$ decreases monotonically from the innermost circular orbit until it becomes zero at the innermost stable circular orbit, the circular time-like geodesic motion can only exist beyond the innermost stable circular orbit. In \mpref{cCase1}, $\lambda_c^2$ in \textbf{Case 1} has a similar behavior as in RN black hole. For the black hole in \textbf{Case 2-1}, the circular geodesic motion exists in three discontinuous range: $2.19387<r<2.84801$, $6.14209<r<6.44301$ and $r>12.4354$. As shown in \mpref{c21a}, the circular motion is always unstable. But in the \mpref{c21b}, there is $\lambda_c^2 < 0$ corresponding to the existence of the stable circular time-like geodesic orbit in the range of $6.14209<r<6.44301$ which is closer to the horizon than in \mpref{c21c}. Compared with RN black hole, the black hole given by \textbf{Case 2-1} has a similar evolution trend of $\lambda_c^2$ in \mpref{c21c} with $r>12.4354$. From \mpref{c21b} and \mpref{c21c}, it seems to be a stable circular orbits closer to the horizon. We guess that this special property of \textbf{Case 2-1} has a Newtonian potential equilibrium position that allows it to have a stable circular time-like geodesic closer to the horizon. For \textbf{Case 2-2} in \mpref{cCase2-2}, the circular orbits are unstable. In \mpref{cCase2-3}, the behavior of $\lambda_c^2$ is similar to that of RN black hole.
\\
\indent Then, we discuss the relationship between the circular motion's Lyapunov exponent $\lambda_c$ and the ``maximal'' Lyapunov exponent $\lambda_s$. In \cite{Hashimoto2017UniversalityIC}, the authors regarded the innermost circular geodesic motion as an equilibrium with a repulsive potential given by particle circular motion. But as shown in \mpref{cRN}, \mpref{cCase1} and \mpref{cCase2-3}, we can see that when the circular time-like geodesic motion approach the innermost circular orbit, $\lambda_s$ will not constraint $\lambda_c$ any more. The ``maximal'' Lyapunov exponent $\lambda_s$ seems to constraint $\lambda_c$ in \mpref{c21a}\footnote{The definition of the ``maximal'' Lyapunov exponent $\lambda_s$ must satisfy the effective potential $V^{''}_{eff}$ has a maximum at the equilibrium position. But for the black hole in Case 2-1, there is no maximum at the equilibrium position in the range $r>4.37690$. See Appendix \ref{upper bound} for detailed discussion.} and \mpref{cCase2-2} with $\lambda_s=\lambda_c$ at the position where Newtonian potential has an equilibrium. We think $\lambda_s$ restricts $\lambda_c$ in \mpref{c21a} and \mpref{cCase2-2} because the particle is close to the equilibrium position of the Newtonian potential where particle motion is slow. In other words, we speculate the main reason for $\lambda_c > \lambda_s$ around the innermost circular orbit in \mpref{cRN}, \mpref{cCase1} and \mpref{cCase2-3} may be when the particle approaches the innermost circular orbit its speed is close to the speed of light. There may be another reason which is the movement in other directions makes Lyapunov exponent growth. To verify our conjecture, we discuss the particle's radial falling in Section \ref{RF}.
\\
\\
\\

\subsubsection{Circular null geodesic and stable photon sphere}
To study the motion of the photon outside the black hole, we set $\eta =0$ in \meqref{rd}. Then, \meqref{rd} can be reduced to
\\
\bea
\dot{r}^2=E^2-\frac{L^2}{r^2}f\left( r\right).\label{cnr}
\eea
\\
From the circular geodesic condition $\dot{r}^2=\left(\dot{r}^2 \right)^{'}=0$, we have constraints on the circular motion of photons
\\
\begin{subequations}
	\begin{align}
	2f\left(r_{cn}\right)&=r_{cn}f^{'}\left(r_{cn}\right),\label{cnc1}
	\\
	\frac{E^2}{L^2}&=\frac{f\left(r_{cn}\right)}{r_{cn}^2},\label{cnc2}
	\end{align}
\end{subequations}
\\
where $r_{cn}$ is the radius of the circular null geodesic, which is defined by \meqref{cnc1}. After taking the black hole solutions we consider \meqref{cnc1} to obtain $r_{cn}$, we substitute the radius $r_{cn}$, \meqref{cnc2} and \meqref{cnr} into \meqref{cgp2} and \meqref{cgc} to calculate the Lyapunov exponent. From the calculation results of the Lyapunov exponent as shown in Table \ref{ps}, we can judge the stability of the circular null geodesics and verify the existence of stable photon spheres.\\
\begin{table}[htpb]
\begin{center}
\begin{tabular}{|c|c|c|c|}
\hline
Black hole&Photon sphere radius&$\lambda^2$ &Stability\\
\hline
\textbf{Case 1}&r=4.72028&$\lambda^2 >0$&unstable\\
\hline
 &r=2.19387&$\lambda^2 >0$&unstable\\
\textbf{Case 2-1}&r=6.44301&$\lambda^2<0$&stable\\
 &r=12.4354&$\lambda^2 >0$&unstable\\
\hline
\textbf{Case 2-2}&r=2.68843&$\lambda^2 >0$&unstable
\\
\hline
\textbf{Case 2-3}&r=14.3266&$\lambda^2 >0$&unstable
\\
\hline
\end{tabular}.
\end{center}
\caption{Photon sphere radius and stability analysis. For the black hole of \textbf{Case 2-1}, there is $\lambda^2< 0$ at $r=6.44301$, which means the photon sphere is stable there.}\label{ps}
\end{table}
\\
\\
\indent From Table \ref{ps}, we can see there is an unstable photon sphere at $r=4.72028$ for \textbf{Case 1}. For \textbf{Case 2-1}, there are three photon spheres at $r=2.19387$, $r=6.44301$, $r=12.4354$, and the one at $r=6.44301$ is stable which agrees with the result in \cite{Liu2019QuasitopologicalED}. For \textbf{Case 2-2} and \textbf{Case 2-3}, the photon spheres are unstable.

\subsection{Radial falling}\label{RF}
In this subsection, we study the geodesic motion of particles falling radially into the black hole. The metric of a 4-dimensional spherically black hole can be written as \meqref{4dbh}. When considering that the particle falls freely towards the black hole in the radial direction, \meqref{rd} can be reduced to
\bea
\dot{r}^2=E^2+\eta f,
\eea
then we can obtain
\bea
\frac{dr}{dt}=\sqrt{f^2+\eta \frac{f^3}{E^2}}.
\eea
When we expand around each point on the particle trajectory (denoted as $r_f$), there is
\bea
\frac{dr}{dt}\sim \frac{2f^{'}+\frac{3\eta}{E^2}ff^{'}}{2\sqrt{1+\eta \frac{f}{E^2}}}\left(r-r_f \right)\label{fme}.
\eea
Similarly, we can get an exponential growth form of the coordinate $r$. Here, we discuss in two cases:
\subsubsection*{massive particle}
For the massive particle falling from infinity, there is time-like geodesic. Setting $\eta=-1, E=1$, we can obtain the Lyapunov exponent $\lambda_{mp}$
 from \meqref{fme}
 \bea
 \lambda_{mp}= \frac{2f^{'}-3ff^{'}}{2\sqrt{1-f}}\label{lmp}.
 \eea
\subsubsection*{photon}
For the photon, there is null geodesic. With $\eta=0$, we have the Lyapunov exponent $\lambda_{ph}$ from \meqref{fme}
\bea
\lambda_{ph}=f^{'}\label{lph}.
\eea
\indent We can see that there is $\lambda_{mp}\approx \lambda_{ph}=2\kappa$ near the horizon\footnote{Near the horizon $r_+$, we have $f=0$ for \meqref{fme}, then \meqref{fme} can be rewritten as $\frac{dr}{dt}\sim f^{'}\left(r-r_+\right)$. Because of the surface gravity $\kappa=\left.\frac{1}{2}f^{'}\right|_{r=r_+}$, we can obtain $r-r_+=e^{2\kappa t}$ at the horizon agreeing with \cite{Zhou2018ParticleMA,wang2019geometry}.}. We study the relationship between these two exponents outside the horizon and the ``maximal'' Lyapunov exponent $\lambda_s$ ($\lambda_s$ can be calculated via \meqref{selambda}).
\\
\indent For simplicity, we first study the RN black hole in \meqref{RN}, then \meqref{lmp}, \meqref{lph} and \meqref{selambda} can be rewritten as
\bea
\begin{aligned}
\lambda_{mp}&=\frac{\left(Q^2-Mr\right)\left(3Q^2-6Mr+r^2\right)}{r^4\sqrt{2Mr-Q^2}},
\quad
\lambda_{ph}&=-\frac{2Q^2}{r^3}+\frac{2M}{r^2},
\quad
\lambda_{s}&=\sqrt{\frac{M^2-Q^2}{r^4}}.
\end{aligned}
\eea
\begin{figure}[htpb]
	\centering
	\subfigure[\scriptsize{}]{
	\includegraphics[scale=0.38]{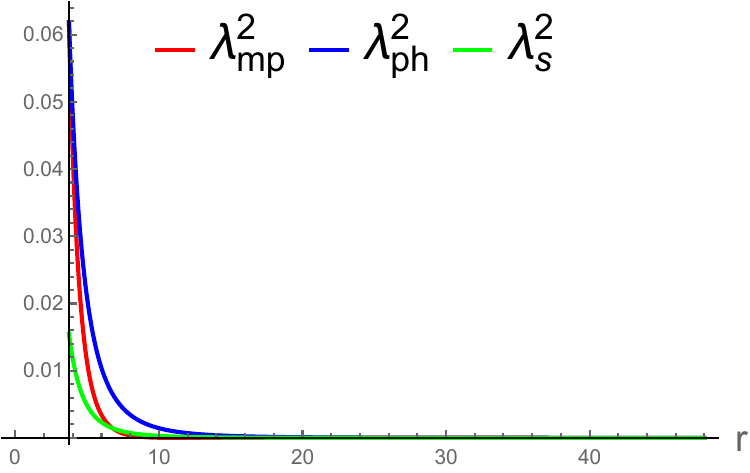}
	}
	\quad
	\subfigure[\scriptsize{}]{
	\includegraphics[scale=0.38]{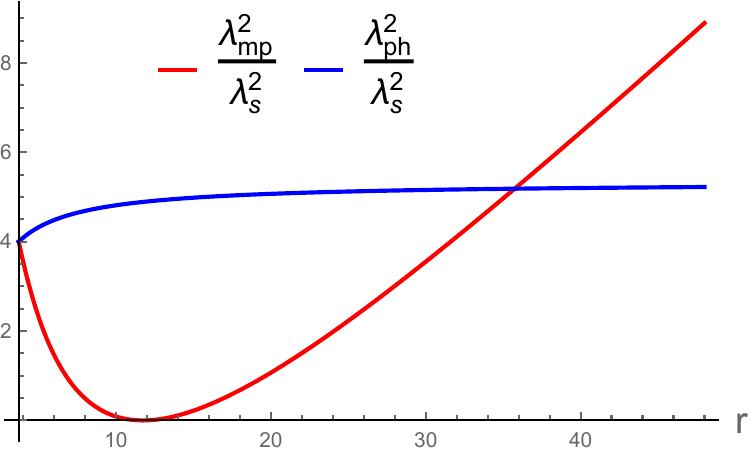}
	}
	\quad
	\subfigure[\scriptsize{}]{
	\includegraphics[scale=0.38]{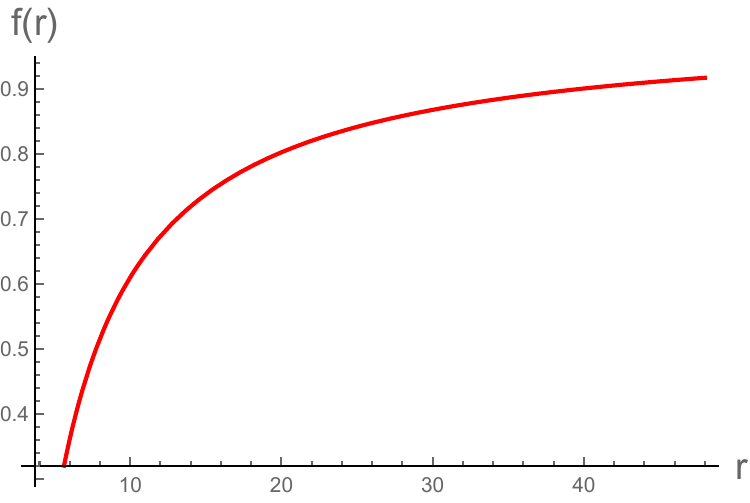}
	}
	\caption{$\lambda_{mp}$, $\lambda_{ph}$ and $\lambda_s$ are the function of $r$. For the RN black hole with $M=2,Q=1$, we can see from (a) that there is $\lambda_{ph}>\lambda_{mp}>\lambda_s$ near the horizon, and $\lambda_{mp}^2$ approaches $\lambda_{ph}^2$. From (b), there is always $\lambda_{ph} \geq 2\lambda_s$.}\label{fRN}
\end{figure}
\\
\indent In \mpref{fRN}, we show the relationship between $\lambda_{mp}$, $\lambda_{ph}$ and $\lambda_{ph}$ in RN black hole and the metric function of RN black hole. There are some interesting phenomena in particle motion near RN black hole. For massive particles, in the region near the horizon, the closer to the horizon, the closer $\lambda_{mp}$ approaches $\lambda_{ph}$. The reason may be the speed of the massive particle approaches the speed of light. For photon which is radial falling into the RN black hole, there is always
\bea
\lambda_{ph} \geq 2\lambda_s\label{f2c}.
\eea
It is similar to $\lambda_{OTOC} \geq 2\lambda_{chaos}$, which is a formula about chaos \cite{Xu2020DoesSE}. Considering the relationship between null geodesic and shock waves, we speculate that there may be some relationship between these two formulas, or even equivalent.
\\
\indent To study the property of \meqref{f2c}, we perform the same calculation for the black holes given by \meqref{case1} and \meqref{case2} and show the results in Figure \ref{fcase1}--\ref{fcase2-3}. From \mpref{fcase1} and \mpref{fcase2-3}, there is always $\lambda_{ph} \geq 2\lambda_s$. But it's not in \mpref{fcase2-1} and \mpref{fcase2-2}. We think the reason is that the metric function $f(r)$ is not monotonically increasing outside the horizon, and we could understand the relationship between $\lambda_{ph} \geq 2\lambda_s$ and $\lambda_{OTOC} \geq 2\lambda_{chaos}$ by studying more examples.
\begin{figure}[htpb]
\centering
\subfigure[\scriptsize{}]{
\includegraphics[scale=0.38]{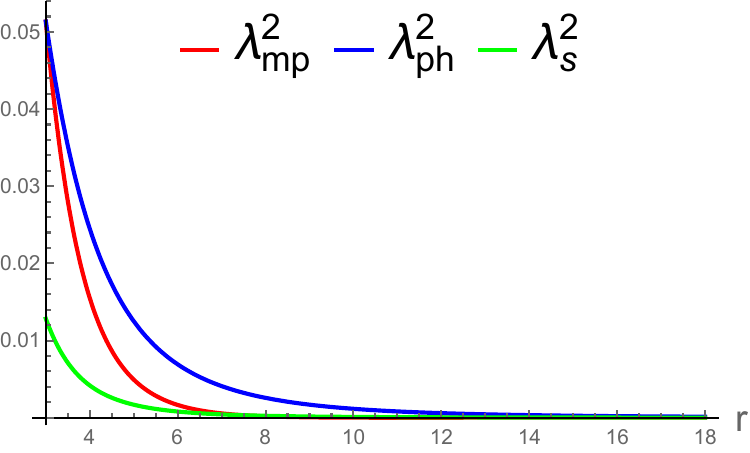}
}
\quad
\subfigure[\scriptsize{}]{
\includegraphics[scale=0.38]{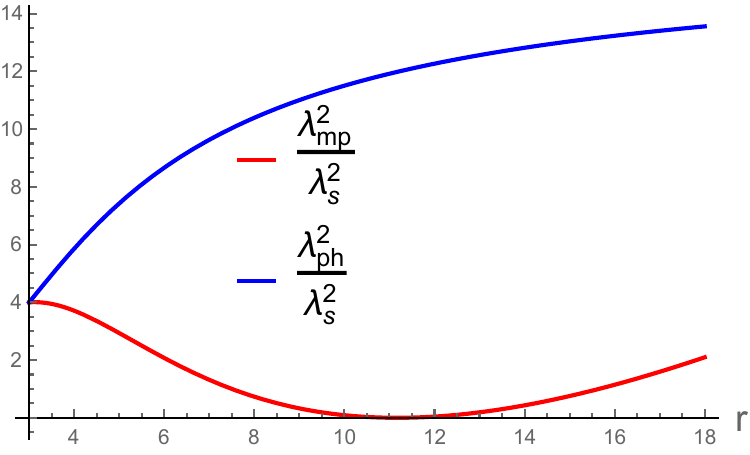}
}
\quad
\subfigure[\scriptsize{}]{
\includegraphics[scale=0.38]{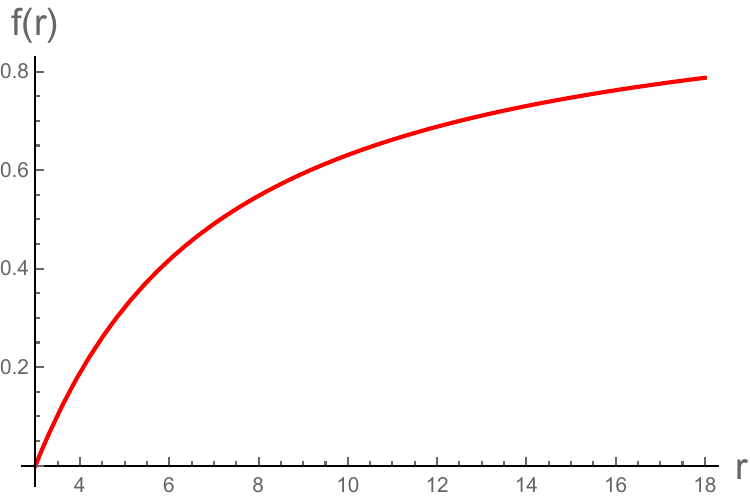}
}
\caption{For the black hole in \textbf{Case 1}, we can see from (a) that $\lambda_{mp}^2$ approaches $\lambda_{ph}^2$ near the horizon. From (b), there is always $\lambda_{ph} \geq 2\lambda_s$ outside the horizon. In(c), the metric function $f\left(r\right)$ increases monotonically.}\label{fcase1}
\end{figure}
\begin{figure}[htpb]
\centering
\subfigure[\scriptsize{}]{
\includegraphics[scale=0.38]{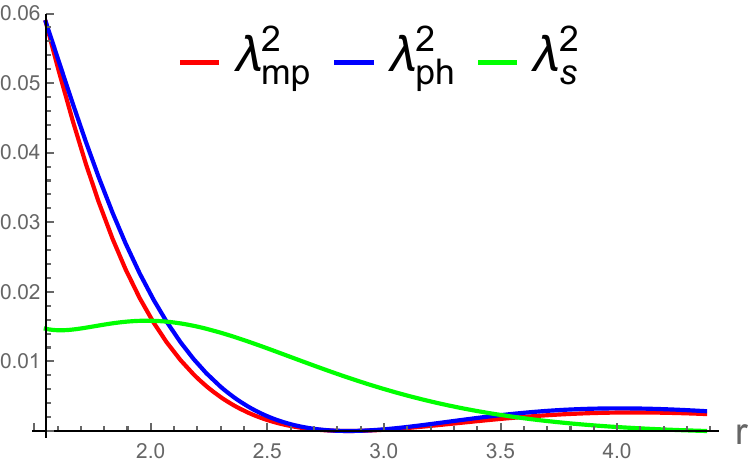}
}
\quad
\subfigure[\scriptsize{}]{
\includegraphics[scale=0.38]{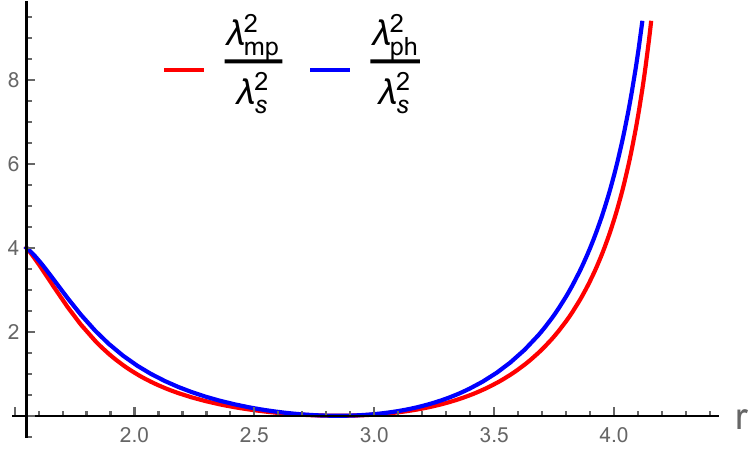}
}
\quad
\subfigure[\scriptsize{}]{
\includegraphics[scale=0.38]{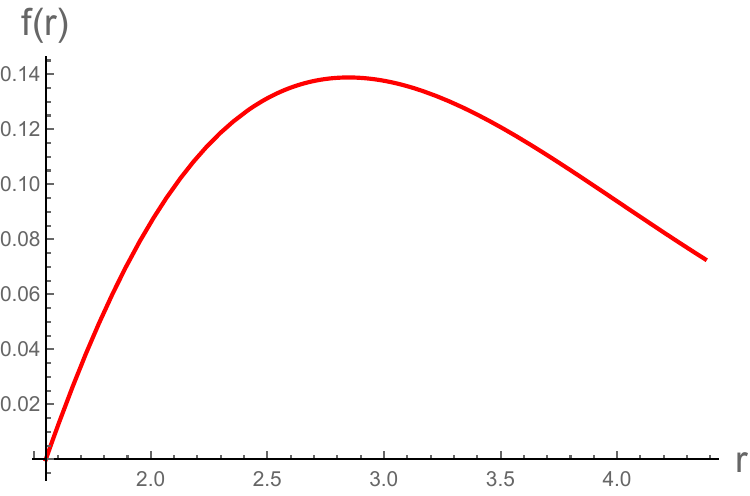}
}
\caption{For the black hole in \textbf{Case 2-1}, $\lambda_{mp}^2$ approaches $\lambda_{ph}^2$ near the horizon in (a). But there is not always $\lambda_{ph} \geq 2\lambda_s$ as shown in (b). Then we observe from (c) that the metric function $f\left(r\right)$ doesn't increase monotonically outside the horizon.}\label{fcase2-1}
\end{figure}
\begin{figure}[htpb]
\centering
\subfigure[\scriptsize{}]{
\includegraphics[scale=0.38]{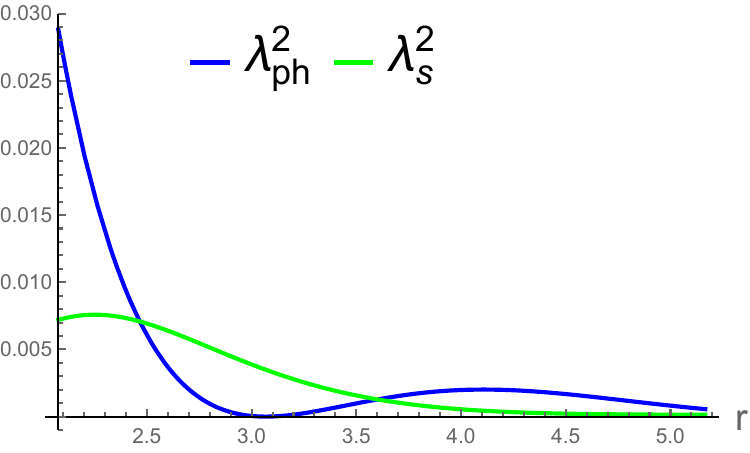}
}
\quad
\subfigure[\scriptsize{}]{
\includegraphics[scale=0.38]{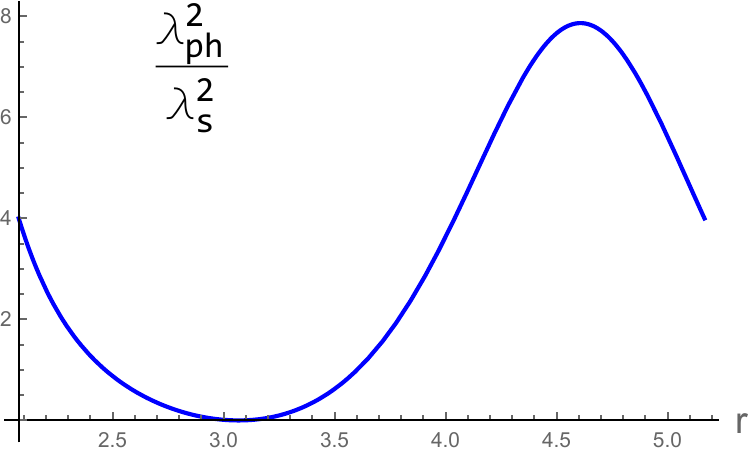}
}
\quad
\subfigure[\scriptsize{}]{
\includegraphics[scale=0.38]{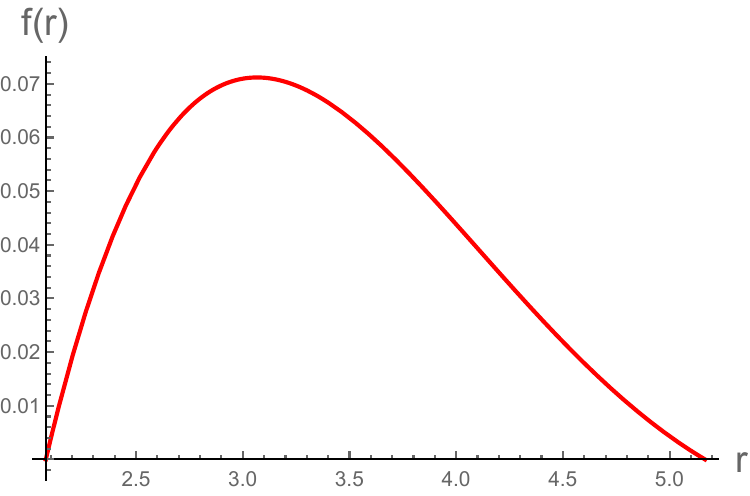}
}
\caption{For the black hole in \textbf{Case 2-2}, there is not always $\lambda_{ph} \geq 2\lambda_s$ in (b). At the same time, the metric function $f\left(r\right)$ doesn't increase monotonically outside the horizon.}\label{fcase2-2}
\end{figure}
\begin{figure}[htpb]
\centering
\subfigure[\scriptsize{}]{
\includegraphics[scale=0.38]{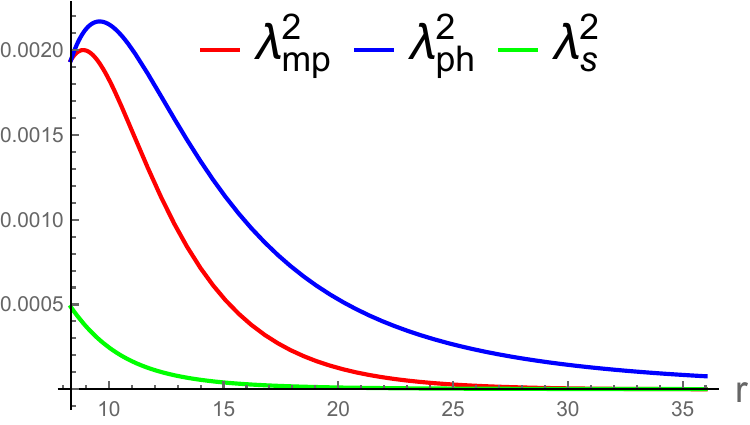}
}
\quad
\subfigure[\scriptsize{}]{
\includegraphics[scale=0.38]{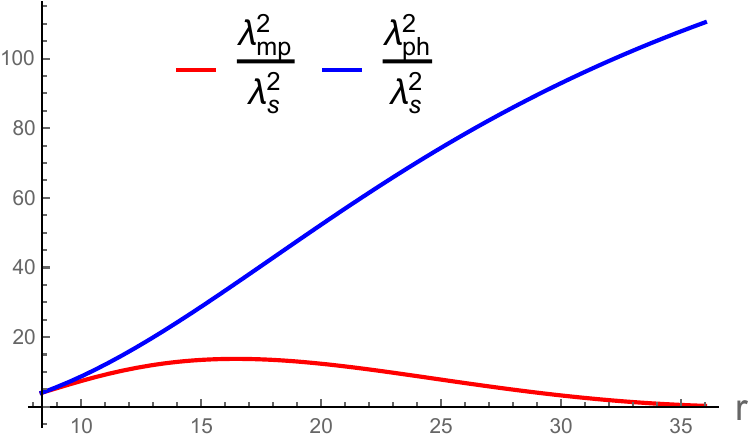}
}
\quad
\subfigure[\scriptsize{}]{
\includegraphics[scale=0.38]{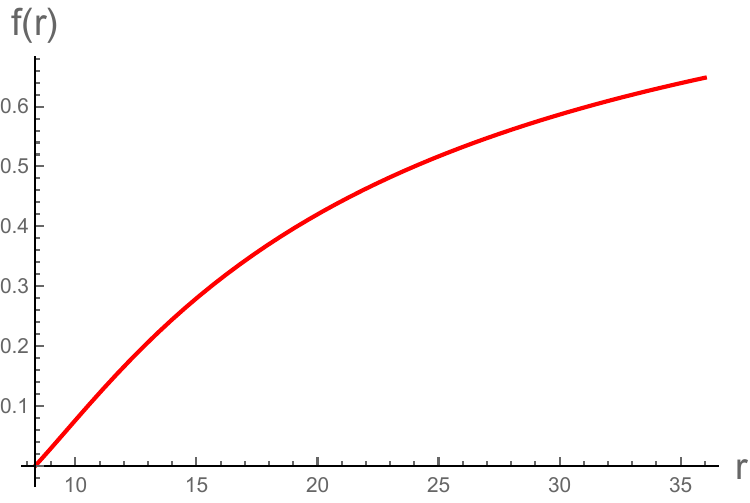}
}
\caption{From the black hole in \textbf{Case 2-3}, we can see from (a) that $\lambda_{mp}^2$ approaches $\lambda_{ph}^2$ near the horizon. $\lambda_{ph} \geq 2\lambda_s$ always exists in (b). Finally, the metric function $f\left(r\right)$ increases monotonically outside the horizon.}\label{fcase2-3}
\end{figure}
\\
\section{Conclusion and discussions}\label{Conclusion and discussions}
In summary, we mainly perform relevant calculations on the black hole solutions given in \cite{Liu2019QuasitopologicalED}. Firstly, we introduce black hole chaos to show why this is an interesting topic. Secondly, we review the definition of $\lambda_s$ which is the ``maximal'' Lyapunov exponent obtained from static equilibrium \cite{Hashimoto2017UniversalityIC, Zhao2018StaticEO}. We calculate the Lyapunov exponent when the charged particle remains static equilibrium near the horizon of these black holes given in Section \ref{A special dyonic black hole}. The results show that the chaos bound does not seem to be universally satisfied. When considering the higher-order expansion term, for black holes with specific parameters, the chaos bound can be violated. Thirdly, we consider a toy model with adjustable parameters to examine the difference between the chaos violated and non-violated cases. From the analysis of Poincar\'e section, we can see the chaos could be strengthened in the system where the chaos bound can be violated, which is a novel study about the black hole chaos.
\\
\indent After studying the static equilibrium of the particles, we turn to the geodesic motion of the particles. We study two types of geodesic motion: circular geodesic motion and radial falling. For circular geodesic motion, we obtain the Lyapunov exponent $\lambda_c$ by the Jacobian matrix and there are some stable circular orbits. At the same time, we find that these Lyapunov exponents $\lambda_c$ of circular geodesic motion do not seem to be constrained by the ``maximal'' Lyapunov exponent $\lambda_s$ defined from static equilibrium. The reason may be that the particle speed approaches the speed of light. This is a topic worth discussing. 
\\
\indent Then we study the particle's radial falling, and find that the Lyapunov exponent $\lambda_{mp}$ of the mass particle will approach the Lyapunov exponent $\lambda_{ph}$ of the photon in the region near the horizon. We speculate the reason may be that the falling speed of the mass particle is close to the speed of light. Note that the discussions are carried out in the geodesic dynamics. It would be interesting to include the backreaction of particle motion to the background spacetime and study the many-body effect of the chaotic behavior. We find that for RN-like black holes or black hole examples where the metric function $f(r)$ increases monotonically outside the horizon, the relation $\lambda_{ph} \geq 2\lambda_s$ is established.
\\
\indent Chaos near black holes is an important subject in contemporary physics research. As far as we know, the nature of the chaos near the black hole will depend on the black hole. Therefore, studying as many different black holes as possible may help us understand chaos more clearly. It may be worth studying the higher-dimensional black hole with quasi-topological electromagnetism given in \cite{cisterna2020quasitopological}.
\\
\section*{Acknowledgement} We would like to thank Hong L$\rm\ddot{u}$, Shu Lin and Qing-Bing Wang for helpful discussions. The work was partially supported by NSFC,
China (grant No.11875184).\\
\\
\section*{Appendix}
\begin{appendix}
\section{The fast Lyapunov indicator (FLI) of the static equilibrium of charged particles}\label{FLI}
\begin{figure}[htp]
\centering
\includegraphics[scale=0.32]{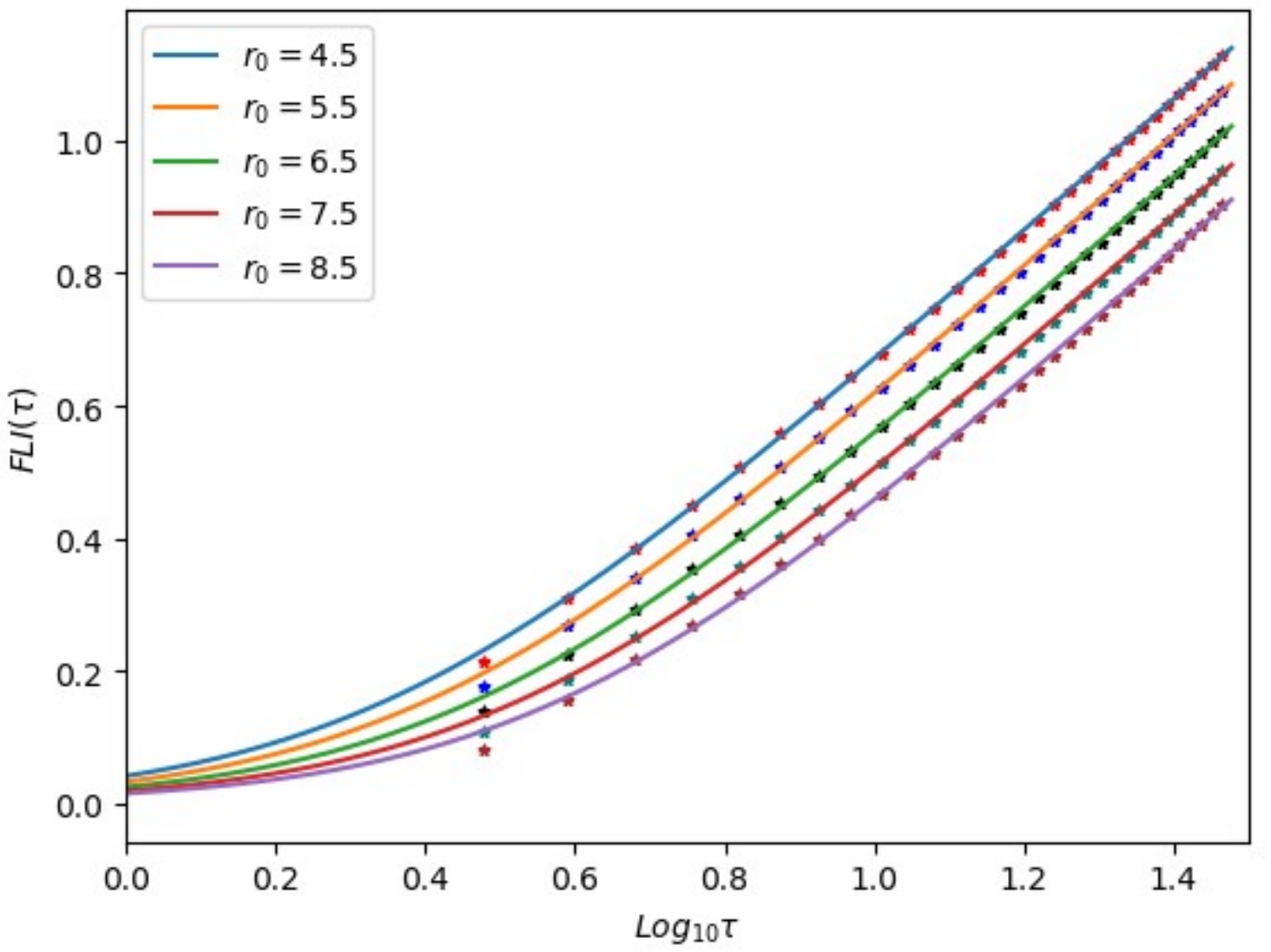}
\caption{Considering the FLI of different equilibrium positions, the equilibrium positions are taking $r_0=4.5,5.5,6.5,7.5,8.5$, and the charge and initial conditions of particles are given accordingly where the perturbation takes $\epsilon=0.00001$. The lines of the different colors are FLI, `` $*$ '' are the exponential fit to the corresponding FLI. All FLIs grow exponentially with $Log_{10}\tau$ at late time, which means chaos is present. The closer the equilibrium position is to the horizon, the faster the FLI grows, which means the greater the chaos.}\label{figFLI}
\end{figure}

 To study the perturbation of particles at static equilibrium more clearly, we use the fast Lyapunov indicator to analyze the perturbation growth. Taking the parameter $k=0$ in \meqref{FLIeq}, we have
\be
FLI(\tau)=\log_{10}\left|\frac{d(\tau)}{d(0)} \right|,
\ee
where $d(\tau)=\sqrt{\left|g_{\mu \nu} \triangle x^\mu \triangle x^\nu\right|}$. $\triangle x^\mu$ is the deviation vector between two nearby trajectories at proper time $\tau$, and in the computation, we choose two particle trajectories with the initial state at the equilibrium position $r_1=r_0$ and the non-equilibrium position $r_2=r_0-\epsilon$ ($\epsilon$ is the perturbation).
\\
\indent We consider a 4d RN black hole metric here
\be
ds^2=-f(r)dt^2+\frac{dr^2}{f(r)}+r^2(d\theta^2+\sin^2\theta d\phi^2),
\ee
where the metric function is $f(r)=1-\frac{2M}{r}+\frac{Q^2}{r^2}$, and the potential function is $A_t(r)=\frac{2Q}{r}$. The trajectory of particles motion can be numerically solved from the equation of motion \meqref{cme}. The parameters of RN black hole here we simply take the mass $M=2$ and the charge $Q=1$. For particle 1, which is initial at the equilibrium position $r_1=r_0$, its initial value condition is $(t,\pi_t,r,\pi_r)=(0,\pi_{t1},r_1,0)$, where $\pi_{t1}=-\sqrt{f(r_1)+f(r_1)^2 \pi_r^2}-q A_t(r_1)$. For particle 2, which is initial at the non-equilibrium position $r_2=r_0-\epsilon$, its initial value condition is $(t,\pi_t,r,\pi_r)=(0,\pi_{t2},r_2,0)$, where $\pi_{t2}=-\sqrt{f(r_2)+f(r_2)^2 \pi_r^2}-q A_t(r_2)$. The charge of two particles are the same, as defined by the equilibrium condition $q=-\frac{(\sqrt{f(r_1)})^{'}}{A_t(r_1)^{'}}$. The results are shown in Figure \ref{figFLI}.
\\
\indent As shown in Figure \ref{figFLI}, for the perturbation on the static equilibrium of charged particles, the chaotic behavior can be found by FLI. The results show that the closer to the horizon, the faster the FLI increases, which is also consistent with the behavior of the ``maximal'' Lyapunov exponent $\lambda_s$ outside the black hole.

\section{Static equilibrium with only gravity}\label{Static equilibrium with only gravity}
\begin{figure}[htp]
\centering
\subfigure[\scriptsize{Case 2-1}]{
\includegraphics[scale=0.6]{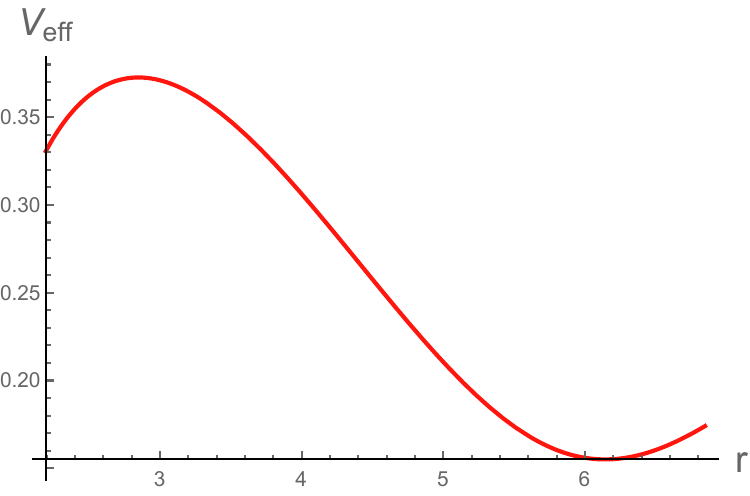}
}
\quad
\subfigure[\scriptsize{Case 2-2}]{
\includegraphics[scale=0.6]{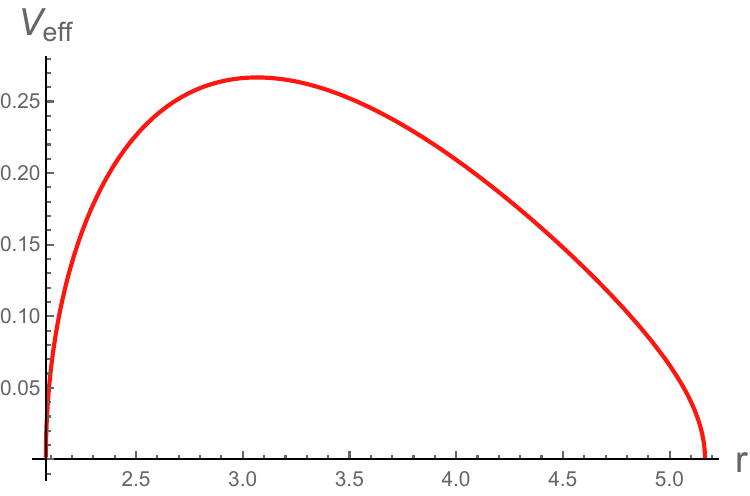}
}
\caption{When only the gravity is considered, for \textbf{Case 2-1}, $V_{eff}$ has the maximum at $r=2.84801$, for \textbf{Case 2-1}, $V_{eff}$ has the maximum at $r=3.06723$.}
\label{ong}
\end{figure}

Since the metric functions of most black holes are monotonically increasing outside the horizon, it is extremely difficult for us to study the situation where particles maintain static equilibrium only under the action of gravitational potential. Fortunately, \textbf{Case 2-1} has an equilibrium position that allows particles to maintain static equilibrium with only gravity.
\\
\indent When only gravitational potential is considered, the effective potential is
\bea
V_{eff}\left(r \right)=\sqrt{f\left(r \right)}\label{gveff},
\eea
and \eqref{slambda} reduces to
\bea
\lambda=\sqrt{-\frac{f\left(r \right) f^{''}\left(r\right)}{2}}\label{nl1}.
\eea
From the analysis of the effective potential curve, we can know that $V_{eff}$ has maximum and minimum values. We will study $\lambda$ at the maximum position of $V_{eff}$ shown in \mpref{ong}. Since this position is far from the black hole horizon, the ``maximal'' Lyapunov exponent $\lambda_s$ should not be given by surface gravity $\kappa$.
\\
\indent We also consider the calculation from the geodesic motion. For the massive particle making a static equilibrium with only gravity, \meqref{rd} can be reduced to
\bea
\dot{r}^2=E^2-f(r).
\eea
With the static condition $\dot{r}^2=\left(\dot{r}^2 \right)^{'}$, we can obtain the coordinate time Lyapunov exponent $\lambda_c$ from \eqref{cgc}
\bea
\lambda_c=\sqrt{-\frac{f(r)f^{''}(r)}{2}}.
\eea
These two calculations have the same result.

\section{The study on static equilibrium by Jacobian matrix}\label{sejm}
When the free particle motion in curved spacetime is calculated, there are two different forms of Lagrangian being often used
\begin{subequations}
\begin{align}
L=&-\sqrt{-g_{\mu \nu}\dot{x}^\mu\dot{x}^\nu}\label{L1},
\\
L=&\frac{1}{2}g_{\mu \nu}\dot{x}^\mu \dot{x}^\nu\label{L2}.
\end{align}
\end{subequations}
\meqref{L1} is defined by the principle of least action, and \meqref{L2} is an equivalent form of \meqref{L1} for simplifying calculation \cite{2019Spacetime}. These two forms are equivalent when calculating particle motion. In \cite{Hashimoto2017UniversalityIC,Zhao2018StaticEO}, the authors obtained the ``maximal'' Lyapunov exponent from the Lagrangian defined by the principle of least action, but the method seems to be unable to the other form of Lagrangian. In this section, we calculate the Lyapunov exponent of the static equilibrium by Jacobian matrix method \cite{cardoso2008geodesic,pradhan2012stability,pradhan2012isco,pradhan2013lyapunov,pradhan2014circular}, which obtains the same results in two form of Lagrangians.
\\
\indent We consider a 4 dimensional spherically symmetric black hole
\be
ds^2=-f(r)dt^2+\frac{dr^2}{f(r)}+r^2(d\theta^2+\sin^2\theta d\phi^2),
\ee
where $f(r)$ is the metric function. The potential function of this black hole is $A_t(r)$. When we discuss the static equilibrium of charged particles, the Lagrangian adds one additional term representing the electric field force compared to \meqref{L1} and \meqref{L2}.
\\
\subsection{The Lagrangian defined by the principle of least action}
The Lagrangian of charged particle near black hole can be write as
\be
L=-\sqrt{-g_{\mu \nu}\dot{x}^\mu\dot{x}^\nu}-\frac{q}{m}\cdot A_t(r)\cdot\dot{t}.
\ee
where q is the charge of particle, m is the mass of particle (taking $m=1$ for simplify the computation), $g_{\mu \nu}$is the metric tensor and $A_t(r)$ is the potential function. With the static gauge $\tau =t$ and the only radial motion, the Lagrangian can be rewritten as
\be
L=-\sqrt{f(r)-\frac{\dot{r}^2}{f(r)}}-q\cdot A_t(r).
\ee
The radial momentum $P_r$ can be defined as
\be
P_r=\frac{\partial L}{\partial \dot{r}}=\frac{\dot{r}/f(r)}{\sqrt{f(r)-\frac{\dot{r}^2}{f(r)}}}.
\ee
\indent From the definition $H=P_r\cdot \dot{r}-L$, the Hamilton can be given as
\be
H=\sqrt{\frac{1+P_r^2\dot f(r)}{f(r)}}+q\cdot A_t(r).
\ee
From the Hamiltonian canonical equation, we can obtain the equation of motion of the particle,
\be
\begin{aligned}
\frac{dr}{dt}&=P_r f(r) \sqrt{\frac{f(r)}{1+P_r^2 f(r)}},
\\
\frac{dP_r}{dt}&=-q\cdot A_t(r)^{'}-\frac{(1+2 P_r^2 f(r))f(r)^{'}}{2\sqrt{f(r)(1+P_r^2 f(r))}}.
\end{aligned}
\ee
For the equilibrium of charged particle, there should be $P_r=\frac{dP_r}{dt}=0$. So we can see at the equilibrium, there is
\be
-q\cdot A_t(r)^{'}-\frac{f(r)^{'}}{2\sqrt{f(r)}}=0.
\ee
There is a static equilibrium condition for the charge of particle at the equilibrium position $r=r_0$,
\be
q=-\left.\frac{(\sqrt{f(r)})^{'}}{A_t(r)^{'}}\right|_{r=r_0}.
\ee
Taken $(r,P_r)$ as the phase space variables, for the dynamic system
\be
\begin{aligned}
\frac{dr}{dt}&=F_1(r,P_r),
\\
\frac{dP_r}{dt}&=F_2(r,P_r),
\end{aligned}
\ee
the components of the Jacobian matrix $K_{ij}$ can be given
\be
\begin{aligned}
K_{11}&=\frac{\partial F_1}{\partial r}=\frac{P_r f(r)^2 \left(2 P_r^2 f(r)+3\right) f^{'}(r)}{2 \left(f(r) \left(P_r^2 f(r)+1\right)\right)^{3/2}},
\\
K_{12}&=\frac{\partial F_1}{\partial P_r}=\frac{f(r)^3}{\left(f(r) \left(P_r^2 f(r)+1\right)\right)^{3/2}},
\\
K_{21}&=\frac{\partial F_2}{\partial r}=\frac{f(r)^{'2}-2 f(r) \left(P_r^2 f(r)+1\right) \left(2 q A_t(r)^{''}(r) \sqrt{f(r) \left(P_r^2 f(r)+1\right)}+\left(2 P_r^2 f(r)+1\right) f(r)^{''}\right)}{4 \left(f(r) \left(P_r^2 f(r)+1\right)\right)^{3/2}},
\\
K_{22}&=\frac{\partial F_2}{\partial P_r}=-\frac{P_r f(r)^2 \left(2 P_r^2 f(r)+3\right) f(r)^{'}}{2 \left(f(r) \left(P_r^2 f(r)+1\right)\right)^{3/2}}.
\end{aligned}
\ee
At the equilibrium location $r=r_0$, where $P_r=\frac{dP_r}{dt}=0$, the Jacobian matrix $K_{ij}$ can be reduced to
\be
K_{ij}=
\left.
\left(
\begin{array}{cc}
0 & f(r)^{3/2} \\
\frac{f(r)^{'2}-2 f(r) \left(2 q \sqrt{f(r)} A_t^{''}(r)+f(r)^{''}\right)}{4 f(r)^{3/2}} & 0 \\
\end{array}
\right)
\right|_{r=r_0}
.
\ee
Substituting $q=-\left.\frac{(\sqrt{f(r)})^{'}}{A_t(r)^{'}}\right|_{r=r_0}$into the Jacobian matrix $K_{ij}$, we can obtain the Lyapunov exponent satisfies
\be
\lambda^2=\left.\frac{1}{4} \left(f(r)^{'2}+2 f(r) \left(\frac{A_t^{''}(r) f(r)^{'}}{A_t'(r)}-f(r)^{''}\right)\right)\right|_{r=r0},
\ee
which is same as \meqref{ls}.
\subsection{The equivalent form of Lagrangian}
Considered another form of the Lagrangian
\be
L=\frac{1}{2}g_{\mu \nu}\dot{x}^\mu \dot{x}^\nu-q A_t \dot{t},\label{c14}
\ee
when only the radial motion is considered, \meqref{c14} can be reduced to
\be
L=\frac{1}{2}\left(-f(r)\dot{t}^2+\frac{\dot{r}^2}{f(r)}\right)-q A_t \dot{t}.
\ee
The generalized momentum can be defined as
\be
\begin{aligned}
\pi_t=&-f \dot{t}-q A_t=-E,
\\
\pi_r=&\frac{\dot{r}}{f}.
\end{aligned}
\ee
Form the definition $H=\frac{1}{2}g^{\mu \nu}(\pi_\mu+q A_\mu)(\pi_\nu+q A_\nu)$, the Hamilton can be given
\be
H=\frac{-(\pi_t+qA_t)^2+\pi_r^2f^2}{2f}.
\ee
Then we can write the equation of motion
\be
\begin{aligned}
\dot{t}=&\frac{\partial H}{\partial\pi_t}=-\frac{q A_t+\pi_t}{f},
\\
\dot{\pi_t}=&-\frac{\partial H}{\partial t}=0,
\\
\dot{r}=&\frac{\partial H}{\partial \pi_r}=f \pi_r,
\\
\dot{\pi_r}=&-\frac{\partial H}{\partial r}=\frac{1}{2}(\frac{2q(\pi_t+q A_t)A_t^{'}}{f}-\pi_r^2 f^{'}-\frac{(\pi_r+q A_t)^2f^{'}}{f^2}).
\end{aligned}\label{cme}
\ee
Rewriting the equation of motion in coordinate time $t$
\be
\begin{aligned}
\frac{dr}{dt}=&\frac{\dot{r}}{\dot{t}}=-\frac{f^2 \pi_r}{q A_t+\pi_t}=F_1,
\\
\frac{d\pi_r}{dt}=&\frac{\dot{\pi_r}}{\dot{t}}=-q A_t^{'}+\frac{(\pi_t+qA_t+\frac{\pi_r^2 f^2}{\pi_t+q A_t})f^{'}}{2 f}=F_2,
\end{aligned}
\ee
we can obtain the components of the Jacobian matrix $K_{ij}$
\be
\begin{aligned}
K_{11}=&\frac{\partial F_1}{\partial r}=\frac{\pi_r f(q f A_t^{'}-2(\pi_t+q A_t)f^{'})}{(\pi_t +q A_t)^2},
\\
K_{12}=&\frac{\partial F_1}{\partial \pi_r}=-\frac{f^2}{\pi_t +qA_t},
\\
K_{21}=&\frac{\partial F_2}{\partial r}=\frac{1}{2}\left(\frac{\pi_r^2f^{'2} }{\pi_t +q A_t}-\frac{(\pi_t +q A_t)f^{'2}}{f^2}-2 q A_t^{''}+\frac{\pi_r^2f(-q A_t^{'}f^{'}+(\pi_t+q A_t)f^{''})}{(\pi_t+q A_t)^2}+\frac{q A_t^{'}f^{'}+(\pi_t+q A_t)f^{''}}{f}\right),
\\
K_{22}=&\frac{\partial F_2}{\partial \pi_r}=\frac{\pi_r f f^{'}}{\pi_t+q A_t}.
\end{aligned}
\ee
Some constraint conditions should be considered. For the motion of charged particle, its orbits should follow the time-like geodesic, which must satisfy the normalization of the four-velocity $g_{\mu \nu}\dot{x}^\mu \dot{x}^\nu=-1$. We can obtain the relation between $\pi_r$ and $\pi_t$
\be
\pi_t=-\sqrt{f+f^2 \pi_r^2}-q A_t.\label{gyh}
\ee
Considering \meqref{gyh} and the equilibrium condition $\pi_r=\frac{d\pi_r}{dt}=0$, the the particle's charge q at the equilibrium position $r=r_0$ should satisfy
\be
q=-\left.\frac{(\sqrt{f(r)})^{'}}{A_t(r)^{'}}\right|_{r=r_0}.
\ee
Then we can see the Lyapunov exponent $\lambda$ should satisfy
\be
\lambda^2=\left.\frac{1}{4} \left(f^{'2}(r)+2 f(r) \left(\frac{A_t^{''}(r) f^{'}(r)}{A_t^{'}(r)}-f^{''}(r)\right)\right)\right|_{r=r_0},
\ee
which is same as \meqref{ls}.

\section{$V_{eff}$ and the ``maximal'' Lyapunov exponent $\lambda_s$}\label{upper bound}
\begin{figure}[htp]
\centering
\subfigure[\scriptsize{Case 1}]{
\includegraphics[scale=0.3]{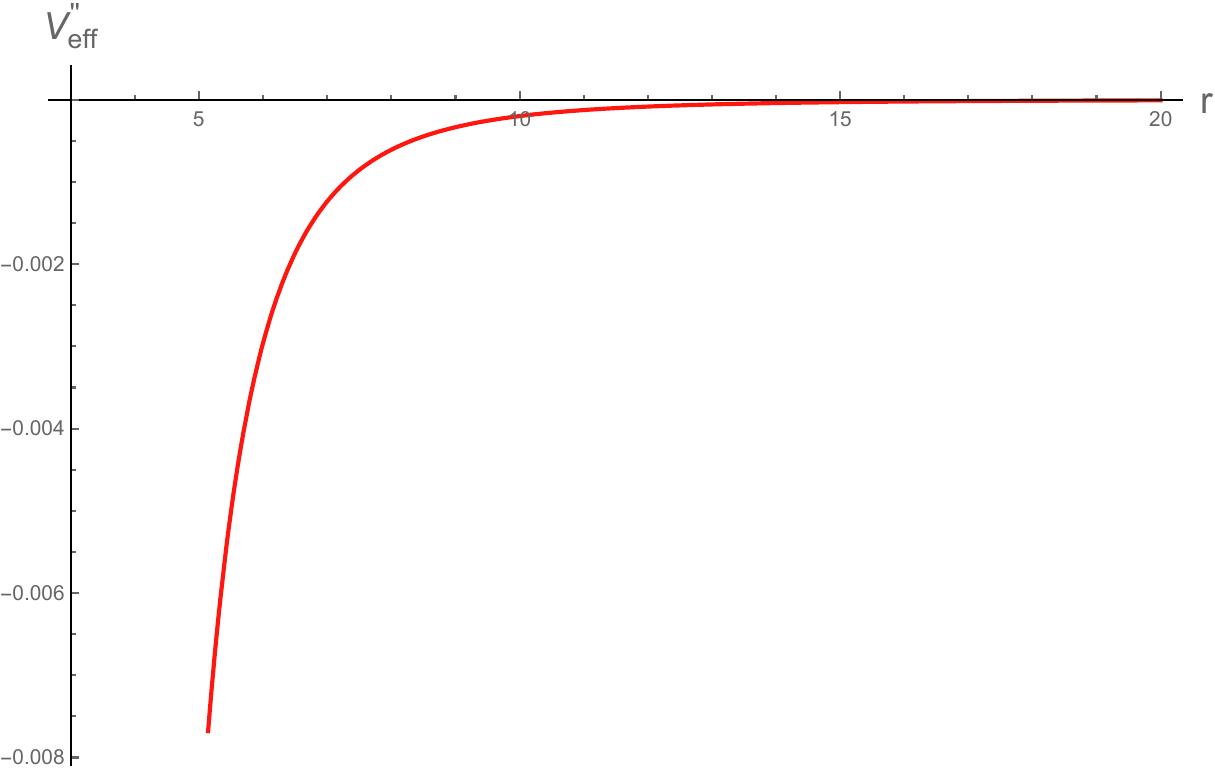}
}
\quad
\subfigure[\scriptsize{Case 2-1}]{
\includegraphics[scale=0.3]{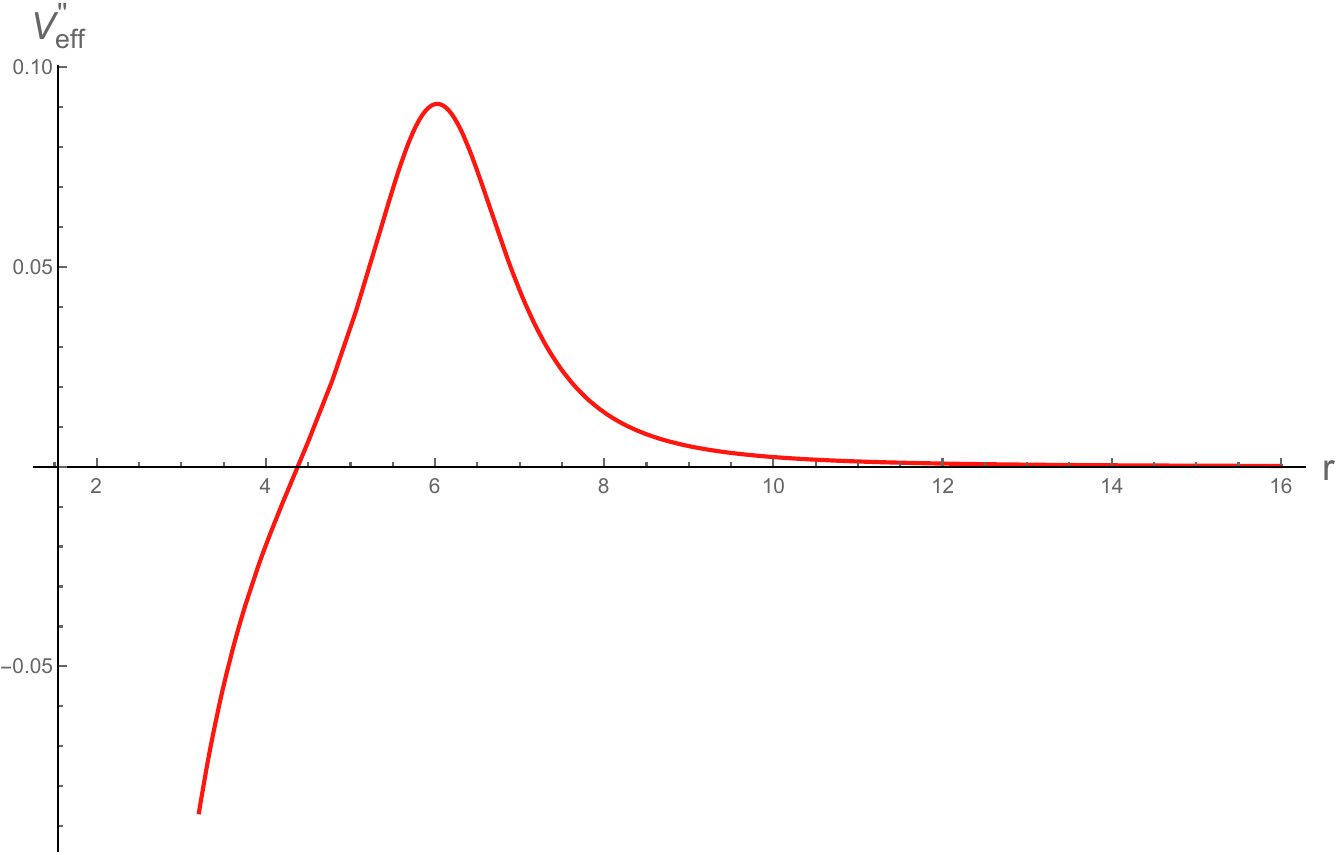}
}
\\
\subfigure[\scriptsize{Case 2-2}]{
\includegraphics[scale=0.42]{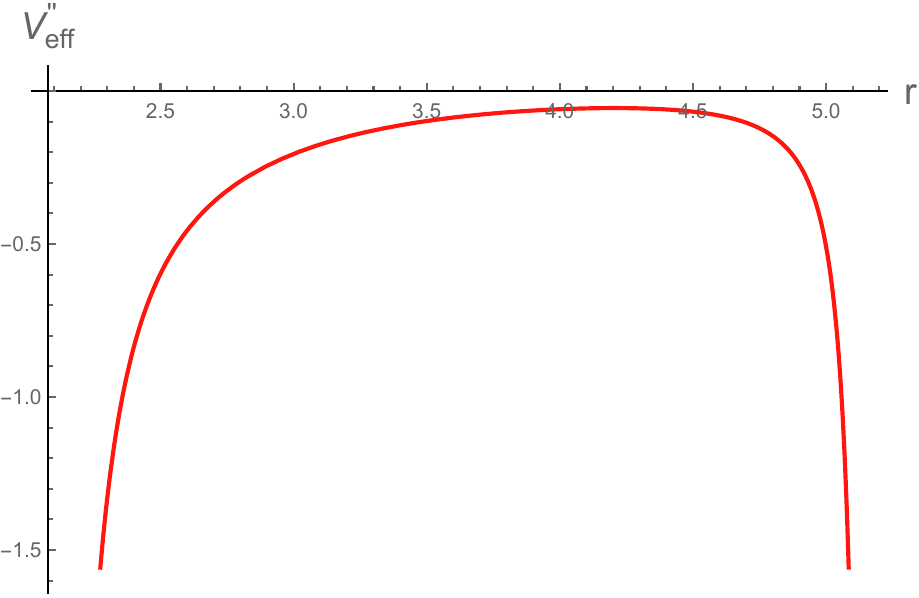}
}
\quad
\subfigure[\scriptsize{Case 2-3}]{
\includegraphics[scale=0.3]{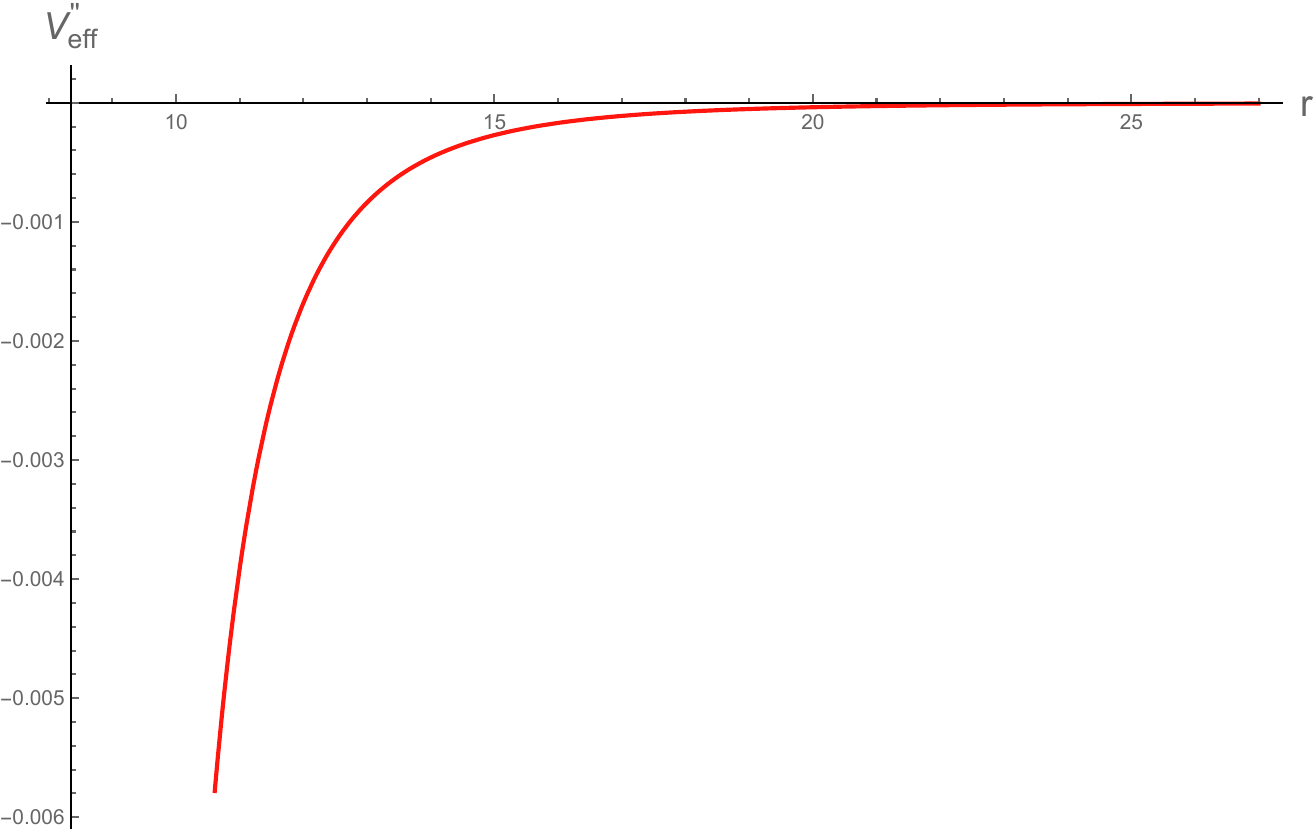}
}
\caption{ $V_{eff}^{''}$ as a function of $r$ corresponds to \meqref{case1} and \meqref{case2}}\label{V2p}
\end{figure}

In studying the static equilibrium of particles to obtain the ``maximal'' Lyapunov exponent $\lambda_s$, the most important point is to ensure that the static equilibrium position is just at the maximum value of the effective potential $V_{eff}$.
\\
\indent For the static equilibrium of charged particles, there is
\\
\bea
V_{eff}=\sqrt{f\left(r\right)}+\frac{e}{m}\Phi_e\left(r\right),
\eea
\\
$V_{eff}^{'}=0$ can be used to determine the static equilibrium position, and on the static equilibrium position, there is
\\
\bea
\frac{e}{m}=-\frac{\left(\sqrt{f\left(r\right)}\right)^{'}}{\Phi_e^{'}\left(r\right)}\label{hzb}.
\eea
\\
When we consider \meqref{hzb}, we can obtain
\\
\bea
V_{eff}^{''}=\left(\sqrt{f\left(r\right)}\right)^{''}-\left(\sqrt{f\left(r\right)}\right)^{'}\frac{\Phi_e^{''}\left(r\right)}{\Phi_e^{'}\left(r\right)}\label{V2}.
\eea
\\
Only when $V^{''}_{eff}$ is negative, the effective potential $V_{eff}$ will have a maximum value at the static equilibrium position, which corresponds to the ``maximal'' Lyapunov exponent $\lambda_s$. As shown in \mpref{V2p}, for the black hole represented by \textbf{Case 2-1}, within a certain range $r>4.37690$, there is always a minimum value for effective potential at the equilibrium position, where the ``maximal'' Lyapunov exponent cannot be defined.

\section{A symplified model}\label{toy model Parameters}
For the toy model we discuss in Section \ref{toy model}, we can study the static equilibrium of particles as we do for the charged particles. The Lagrangian of particles which move in $x$ direction($y=\dot{y}=0$) can be written as
\be
\mathcal{L}=- \sqrt{f{\left (x \right )}- \frac{\dot{x}^{2}}{f{\left (x \right )}} - \dot{y}^{2} + } - \omega A(x),
\ee
where ``$\cdot$'' denotes derivative with respect to the coordinate time $t$. We can obtain the Lyapunov exponent $\lambda$ at the equilibrium position $x=x_0$, $\lambda$ satisfies
\be
\lambda^2=\left.\frac{1}{4} \left(f^{'2}(x)+2 f(x) \left(\frac{A^{''}(x) f^{'}(x)}{A^{'}(x)}-f^{''}(x)\right)\right)\right|_{x=x_0}.\label{xlambda}
\ee
\indent When we consider the near-horizon behavior of particle, \meqref{xlambda} can be expanded at the horizon $x=x_h$ as
\be
\lambda^2=\kappa^2+\gamma(x-x_h),
\ee
where $\kappa=\frac{f_1}{2}$ and $\gamma=4\kappa^2\frac{A^{''}(x_h)}{A^{'}(x_h)}$. If $\gamma>0$, there is violation of chaos bound, or the chaos bound is not violated when $\gamma \leq 0$. With $A(x)=a(x-x_c)^2+b(x-x_c)^4$, $\gamma$ can be rewritten as
\be
\gamma=4\kappa^2 \frac{2 a + 12 b \left(- x_{c} + x_{h}\right)^{2}}{a \left(- 2 x_{c} + 2 x_{h}\right) + 4 b \left(- x_{c} + x_{h}\right)^{3}}.\label{gammacondition}
\ee
We set $x_h=0,x_c=1$ as in \cite{Hashimoto2017UniversalityIC}, then $\gamma$ is positive or not can be decided by adjusting the values of $a$ and $b$. Moreover, in order to avoid particles falling into horizon, we study the Hamiltonian of particles which satisfy $x=x_h,\ ,y=0,\ \frac{dx}{dt}=0$
\be
H=\omega(a+b).
\ee
For simplifying the calculation, we add an additional condition $a+b=1$. There is $E_{max}=\omega$, for the particles whose energy $E$ satisfies $E<E_{max}$, they will not fall into horizon.
\\
\indent From \meqref{gammacondition} and the parameters of $x_c$ and $x_h$, we can find the conditions for parameters corresponding to the chaos bound can be violated ($\gamma >0$) and the chaos bound cannot be violated ($\gamma \leq 0$):
\\
\\
\emph{1. Chaos bound can be violated ($\gamma >0$) case}
\begin{equation*}
\frac{a+6b}{-2a-4b} >0.
\end{equation*}
One of its solutions is $-2b<a<-6b$ while $a>0$ and $b<0$. With $a+b=1$, we can set $a=1.3, \quad b=-0.3$.
\\
\emph{2. Chaos bound not violated ($\gamma \leq 0$) case}
\begin{equation*}
\frac{a+6b}{-2a-4b} <0.
\end{equation*}
There is a solution $a>-6b$ while $a>0$ and $b<0$. With $a+b=1$, we can set $a=1.1, \quad b=-0.1$.

\end{appendix}

\end{document}